\documentclass[sn-mathphys-num]{sn-jnl}% Math and Physical Sciences Numbered Reference Style 
\usepackage{graphicx}%
\usepackage{multirow}%
\usepackage{amsmath,amssymb,amsfonts}%
\usepackage{amsthm}%
\usepackage{mathrsfs}%
\usepackage[title]{appendix}%
\usepackage{xcolor}%
\usepackage{textcomp}%
\usepackage{manyfoot}%
\usepackage{booktabs}%
\usepackage{algorithm}%
\usepackage{algorithmicx}%
\usepackage{algpseudocode}%
\usepackage{listings}%
\usepackage{natbib}
\usepackage{bm}
\theoremstyle{thmstyleone}%
\theoremstyle{thmstyletwo}%

\theoremstyle{thmstylethree}%

\begin{document}

\title[Article Title]{Numerical investigation of wake dynamics and heat transfer in MHD flows around confined triangular prisms}

%%=============================================================%%
%% GivenName	-> \fnm{Joergen W.}
%% Particle	-> \spfx{van der} -> surname prefix
%% FamilyName	-> \sur{Ploeg}
%% Suffix	-> \sfx{IV}
%% \author*[1,2]{\fnm{Joergen W.} \spfx{van der} \sur{Ploeg} 
%%  \sfx{IV}}\email{iauthor@gmail.com}
%%=============================================================%%

\author[1]{\fnm{Amulya Sai } \sur{Akkaladevi}}

\author[1]{\fnm{Prabhat} \sur{Kumar}}
%\equalcont{These authors contributed equally to this work.}

\author*[1]{\fnm{Sachidananda} \sur{Behera}}\email{sbehera@mae.iith.ac.in}
%\equalcont{These authors contributed equally to this work.}

\affil[1]{\orgdiv{Department of Mechanical and Aerospace Engineering}, \orgname{Indian Institute of Technology Hyderabad}, \orgaddress{\street{Kandi}, \city{Sangareddy}, \postcode{502285}, \state{Telangana}, \country{India}}}

%%==================================%%
%% Sample for unstructured abstract %%
%%==================================%%

\abstract{This study numerically investigates the flow evolution and heat transfer characteristics of an electrically conducting fluid over triangular prisms confined between two parallel plates with a heated bottom plate under the influence of a magnetic field. The research focuses on the three-dimensional behavior of MHD flows at low Hartmann numbers ($Ha$), exploring how obstacle orientation and mixed convection influence flow dynamics and heat transfer. Three-dimensional simulations are performed using an in-house MHD solver in OpenFOAM at a constant channel height based Reynolds number ($Re_{h}=600$). The combined effects of Richardson number ($Ri$) and $Ha$ on wake dynamics and heat transfer are analyzed for three triangular prism orientations. The results reveal that increasing $Ha$ promotes flow two-dimensionality, while higher $Ri$ enhances three-dimensionality. Three wake instability modes (Mode A, B, and C) are identified. Orientation 2 exhibits the lowest mean drag coefficient at $Ha=0$, $Ri=5$, while the highest mean lift coefficient is observed at $Ha=0$, $Ri=0$. Orientation 3 achieves the highest heat transfer rate, with an average Nusselt number of $21.05$ at $Ha=25$, $Ri=5$, and consistently outperforms the other orientations in heat transfer across various $Ha$ and $Ri$ conditions. These findings highlight the strong coupling between wake dynamics and heat transfer, offering insights for optimizing MHD flows in practical applications.}


%%\pacs[JEL Classification]{D8, H51}

%%\pacs[MSC Classification]{35A01, 65L10, 65L12, 65L20, 65L70}

\maketitle

\section{Introduction}\label{sec1}

Magnetohydrodynamics (MHD) refers to the study of the flow behavior of electrically conducting fluids in the presence of a magnetic field. MHD plays a crucial role in various industrial applications, including cooling of nuclear reactors, electromagnetic stirring in casting processes, plasma confinement, etc. It is also an important area of research in various natural phenomena like astrophysics and geophysics. Liquid metals are commonly used in industrial MHD applications due to their high thermal conductivity, low kinematic viscosity, low vapor pressure, and ability to stay liquid at high temperatures \citep{ching1981laminar}. These properties make liquid metals ideal coolants for advanced nuclear reactor cooling systems \citep{gajbhiye2018validation} and suitable working fluids in heat exchangers. When liquid metal flows through a confined domain under an applied magnetic field, it induces electric potential and current \citep{gajbhiye2015numerical}. The interaction of the induced current with the magnetic field generates an electromagnetic force, known as the Lorentz force, which alters or suppresses the flow within the confined domain \citep{ozoe1989effect,davidson2002introduction,yoon2004numerical, kanaris2013three}.

Enhancing heat transfer is vital in MHD applications such as nuclear reactor cooling and heat exchangers, where efficient thermal management is critical. The use of bluff bodies can introduce turbulence, which promotes better mixing within the flow, leading to a significant improvement in heat transfer performance. Several experimental \citep{josserand1993pressure, lahjomri1993cylinder, kim2000investigation} and numerical \citep{mutschke1998cylinder, muck2000three, mutschke2001scenario, dousset2008numerical, grigoriadis2010mhd, kanaris2013three} studies have been conducted in the past to explore the complex two and three-dimensional wake flow dynamics for various obstacle shapes under the influence of external magnetic fields. For instance, in a 2D numerical study on the flow past a circular cylinder under an aligned magnetic field, \citet{yoon2004numerical} observed that as the Hartmann number ($Ha$) increases, vortex shedding is suppressed, reducing the lift coefficient and resulting in a steady flow with two narrow and elongated eddies. \citet{kanaris2013three} showed that the suppression of vortices depends on their alignment with the magnetic field. Vortices parallel to the field are mainly influenced by Hartmann damping, whereas those at an angle are strongly suppressed by the Lorentz force, emphasizing the importance of magnetic field orientation in flow behavior. This relationship between the magnetic field and flow dynamics is further quantified by \citet{sekhar2006flow}, who reported that the aerodynamic drag coefficient is proportional to $N^{0.5}$, where $N = \frac{ \sigma LB^{2}}{\rho U}$ represents the ratio of electromagnetic to inertial forces. Similarly, \citet{chatterjee2012control}, in their study on liquid metal flow past a circular cylinder, observed that the separation bubble length behind the cylinder gradually decreases and becomes obsolete at a critical Hartmann number.

While much of the discussion so far has centered on flow dynamics, considerable research has also been devoted to investigating heat transfer characteristics in MHD flows, particularly in the context of obstacles and the influence of magnetic fields. \citet{hussam2011dynamics} and \citet{hussam2013heat} investigated MHD duct flow with a circular cylinder and found that the heat transfer is strongly influenced by the proximity of the cylinder to the heated wall. They also reported a significant increase in the Nusselt number as the blockage ratio ($\beta$) is increased at a constant Hartmann number ($Ha$). Later on, \citet{hussam2018effect} extended their work by performing a numerical analysis to examine heat transfer enhancement in liquid metal flow using triangular and square-shaped obstacles. They found that at low magnetic fields ($Ha \le 320$), square obstacles provided higher heat transfer enhancement ($78\%$) compared to triangular obstacles ($75\%$). However, at a high magnetic field ($Ha = 2400$), triangular obstacles were more effective in perturbing the flow and improving heat transfer than square obstacles. Similarly, \citet{chatterjee2013wall, chatterjee2015mhd} conducted numerical analyses to study flow perturbation caused by circular and square-shaped cylinders in the presence of a magnetic field. Their findings indicated that placing a square cylinder in the channel enhanced heat transfer compared to a bare channel. However, replacing the square cylinder with a circular one of the same size resulted in a reduction in heat transfer. \citet{farahi2017investigation} performed quasi-two-dimensional simulations of MHD flow around a circular cylinder, and reported a $48\%$ increase in heat transfer when the circular cylinder was offset from the wake centerline. In a recent study, \citet{singh2019numerical} performed the numerical simulations of MHD  mixed convection flow over a diamond shaped obstacle. They concluded that the heat transfer decreases with an increase in $Ha$, and conversely, the heat transfer increases with buoyancy forces.

Most prior research on two-dimensional MHD flows \citep{potherat2000effective,frank2001visual,hussam2011dynamics,dousset2008numerical} has concentrated on high Hartmann numbers ($Ha>320$), where strong Lorentz forces suppress transverse velocity components, resulting in quasi-two-dimensional flow behavior. Although these studies have significantly advanced our understanding of the flow dynamics of MHD at high $Ha$, they overlook the intricate three-dimensional flow phenomena that emerge at lower $Ha$, where electromagnetic forces are weaker and inertial effects dominate. Moreover, most of these studies focus on symmetric geometries such as circular and square cylinders, which, while providing valuable insights, fail to capture the complex flow and thermal behaviors associated with asymmetric geometries such as triangular prisms. Unlike symmetric shapes, triangular prisms introduce unique wake structures and boundary layer separation patterns, which are highly sensitive to their orientation relative to the flow direction. These variations can profoundly impact vortex formation, wake instabilities, and heat transfer performance, yet they remain largely unexplored. Additionally, while buoyancy-induced convection and electromagnetic forces are individually well documented, their interaction in MHD flows involving asymmetric bluff bodies has not been systematically studied. Buoyancy, particularly when acting perpendicular to the main flow, introduces a competing mechanism that promotes three-dimensionality, counteracting the two-dimensional tendencies induced by Lorentz forces. This interplay adds a layer of complexity that has been overlooked in the existing literature, leaving a significant gap in our understanding of the physics governing such flows.

The present study aims to address several critical questions to advance the understanding of three-dimensional magnetohydrodynamic (MHD) flows. Specifically, it seeks to: (i) investigate how three-dimensional instabilities develop under the combined effects of magnetic fields and buoyancy forces, particularly at low Hartmann numbers ($Ha$); (ii) examine the complex interplay between buoyancy-driven convection and Lorentz forces and their combined influence on wake dynamics and heat transfer in confined domains; and (iii) assess the impact of triangular prism orientation on wake behavior, flow separation, and thermal performance, identifying configurations that optimize heat transfer efficiency. By addressing these questions, the study provides new insights into the fundamental physics of MHD flows and practical strategies for improving thermal management systems in industrial applications. To address these questions, this study conducts three-dimensional simulations of MHD flows over triangular prisms confined between parallel plates with a heated bottom wall. Using an in-house developed MHD solver in OpenFOAM, simulations are performed at a fixed Reynolds number ($Re_{h}=600$) for varying $Ha$ and $Ri$, considering three distinct prism orientations. The results provide insights into wake dynamics, recirculation length, drag and lift forces, and heat transfer performance, bridging critical knowledge gaps in MHD flow studies.

The rest of the paper is organized as follows. Subsection \ref{ssec:gv} discusses the governing equations and the important non-dimensional parameters, and $\S$ \ref{ssec:nm} describes the numerical method used in the present study. The details of the simulation set-up are presented in $\S$ \ref{ssec:setup}. The code validation and grid-independence studies are presented in $\S$ \ref{sec:valgrid}. The simulation results on the effect of the strength of the applied magnetic field and mixed convection effects on the wake behavior behind the cylinder and the associated heat transfer characteristics are discussed in $\S$ \ref{sec:reslt_dis}. Finally, $\S$ \ref{sec:concl} concludes the key findings of the present study.

\section{Simulation methods  \label{sec:soln}}
\subsection{Governing equations \label{ssec:gv}}
The three-dimensional, incompressible Navier-Stokes equations governing the flow of a Newtonian, electrically conducting fluid in the presence of a magnetic field, accounting for the Lorentz force \citep{GAJBHIYE2018168,singh2019numerical_,grigoriadis2010mhd}, are expressed as:
\begin{equation}
	\nabla \cdot \mathbf{u} =0
	\label{Eq1}
\end{equation}
\begin{equation}
	\frac{\partial \mathbf{u}}{\partial {t}} + (\mathbf{u} \cdot \nabla)\mathbf{u}= -\frac{\nabla {p}}{\rho} + \nu (\nabla^{2} \mathbf{u}) + \frac{(\mathbf{j} \times \mathbf{B})}{\rho} - \beta (T-T_{0})\mathbf{g}
	\label{Eq2}
\end{equation}

The heat transfer resulting from mixed convection is accounted for by coupling the incompressible Navier-Stokes equation with the energy equation \citep{singh2019numerical,singh2019numerical_,GAJBHIYE2018168}, which is expressed as follows:
\begin{equation}
	\frac{\partial {T}}{\partial {t}} + (\mathbf{u} \cdot \nabla)T=  \alpha (\nabla^{2} {T})
	\label{Eq3}
\end{equation}
For the range of Rayleigh numbers examined in this study, the Boussinesq approximation is deemed valid. Consequently, the buoyancy force is incorporated as a source term in the Navier-Stokes equation. Additionally, the magnetic Reynolds number is assumed to be negligible ($Re_m << 1$), as the induced magnetic fields are considerably smaller compared to the externally applied magnetic field. To incorporate magneto-hydrodynamics effects in the flow, the electric potential equation is solved alongside the momentum and energy equations and is represented as follows:
\begin{equation}
	\textbf{j}= \sigma \left ( -\nabla {\phi}+ \left [ \mathbf{u} \times \mathbf{B} \right ] \right )
	\label{Eq4}
\end{equation}
\begin{equation}
	\nabla \cdot \textbf{j}=0
	\label{Eq5}
\end{equation}
The present study focuses on several key parameters that influence the flow behavior, and these parameters are expressed as dimensionless numbers. The Hartman number ($Ha$) represents the relative importance of magnetic forces compared to viscous forces acting on the fluid. The Richardson number ($Ri$) indicates the balance between buoyancy-driven flow (also known as free convection) and forced flow. The Reynolds number ($Re$) reflects the ratio of inertial forces to viscous forces. Finally, the Prandtl number ($Pr$) compares the rate of momentum diffusion to thermal diffusion within the fluid. The specific definitions of these dimensionless numbers are provided below.
\begin{align*}
	Ha=Bh\sqrt{\frac{\sigma}{\rho\nu}}&& 
	Re=\frac{Uh}{\nu}&& Gr= \frac{g\beta\Delta T h^{3}}{\nu^{2}}&& Ri=\frac{Gr}{Re^2}&&  Pr= \frac{\nu}{\alpha}
\end{align*}
In the equations (\ref{Eq1})--(\ref{Eq5}), \textbf{u}, p, T, \textbf{j}, \textbf{B}, and $\phi$, denotes the velocity vector, pressure field, temperature field, current density vector, magnetic field vector, and electric potential, respectively. The parameters $\rho$, $\nu$, $\beta$, $\sigma$, and $\alpha$ represent the fluid properties such as density, kinematic viscosity, thermal expansion coefficient, electrical conductivity, and thermal diffusivity, while $g$ represents the acceleration due to gravity. 

\subsection{Numerical methods \label{ssec:nm}}
The governing equations (\ref{Eq1})--(\ref{Eq5}) are solved using an unsteady incompressible flow solver in OpenFOAM, an open-source CFD code. OpenFOAM uses the finite volume method on a collocated grid arrangement for the discretization of the governing equations. The simulations are performed using the PIMPLE algorithm, a native solver to OpenFOAM, which combines aspects of PISO and SIMPLE algorithms. The PIMPLE algorithm utilizes the Rhie Chow momentum interpolation methods to ensure pressure velocity coupling. The pressure Poisson equation, which couples the pressure and velocity field, is solved using the conjugate gradient method to determine the pressure field. The computational domain is discretized using a Cartesian body-fitted non-uniform grid system, where the grids are denser near solid walls. The minimum grid size, $\Delta_{min}$, near the walls, is kept as $\Delta_{min}=0.0045h$, and the maximum grid size, $\Delta_{max}$, in the computational domain is kept as $\Delta_{max}=0.075h$, for all the simulations. This approach ensures efficient use of computational resources while maintaining accuracy in critical areas. For spatio-temporal accuracy of the numerical solution, the simulations employed second-order accurate discretization schemes for space derivatives and first-order accurate schemes for time derivatives. Additionally, a Courant number (Co) strictly less than one ($Co < 1$) is maintained throughout all simulations to ensure stability. For code validation and grid-independent solutions, comprehensive validation and grid-independence studies are conducted, and the details are provided in subsection~\ref{sec:valgrid}.

\subsection{Simulation set-up \label{ssec:setup}}
\begin{figure}
	\centering
	% \begin{subfigure}
	%      \centering
	%      \includegraphics[width=1\textwidth]{Fig_1a.eps}
	% \label{fig:enter-label}
	%  \end{subfigure}   
	%  \begin{subfigure}
	%      \centering
	%      \hspace{0.6cm} {\large ($a$)} \hspace{3cm} {\large ($b$)} \hspace{2.8cm} {\large ($c$)} 
	%      \includegraphics[width=1\textwidth]{Fig_1b.eps}
	%  \end{subfigure} \\
	% \hspace{1.5cm} {\large ($a$)} \hspace{3.5cm} {\large ($b$)} \hspace{3.5cm} {\large ($c$)}     
	\includegraphics[width=1\textwidth]{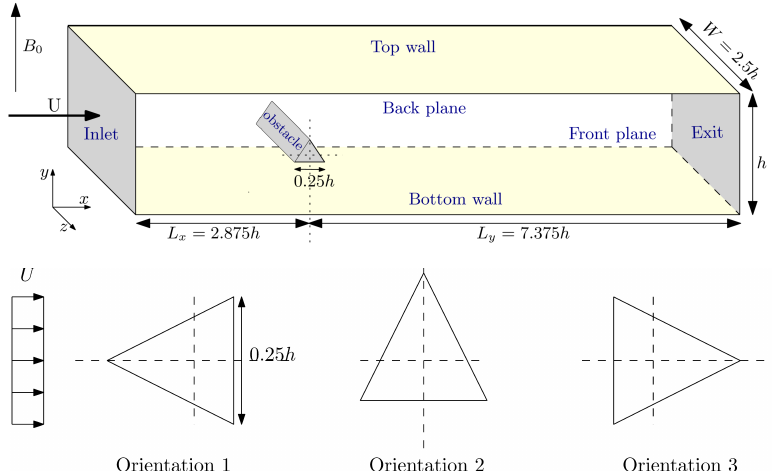}
	\caption{Three-dimensional computational domain featuring dimensions scaled relative to characteristic length ($h$). The prism in focus is explored in three distinct orientations. These orientations are depicted in front views:  (a) Orientation 1, with the base perpendicular to the flow direction (b) Orientation 2, where the base aligns with the flow direction and the prism is rotated by 30 degrees about its centroid, and (c) Orientation 3, rotated by 60 degrees about its centroid. The magnetic field is applied in the vertical direction normal to the flow.}
	\label{orientations}
\end{figure}

The three-dimensional computational domain for the present study is shown in Fig.~\ref{orientations}. The simulations are performed in a rectangular duct with a height of ($h$) and a length of ($10.25h$). To mimic flow between parallel plates, the front and back faces are specified as free-slip boundaries, while the top and bottom walls are considered as no-slip walls. An equilateral triangular prism with a side length of ($0.25h$) is placed inside the duct, with its center located ($2.875h$) downstream from the inlet. The magnetic field is applied in the vertical direction ($y$-axis). The width of the domain in the $z$-direction is taken as ($2.5h$). At the inlet to the computational domain, a uniform velocity field ($u=U, v=0, w=0$) is specified, while an outflow boundary condition is applied at the exit. The bottom wall of the channel acts as a hot wall, maintained at a constant temperature of ($T_{h}$). The inlet flow and the top wall are kept at a constant temperature of ($T_{c}$) lower than the top wall. All other boundaries, including the triangular prism itself, are considered thermally insulated. The prism is also considered to be electrically insulated. The working fluid is assumed to be incompressible, viscous, and electrically conducting. The fluid properties are considered constant and independent of temperature and magnetic field variations. A total of $27$ simulations were conducted, exploring various combinations of Richardson numbers ($Ri = 0, 1,$ and $5$) and Hartmann numbers ($Ha = 0, 25,$ and $50$) for each prism orientation. The Prandtl number for the fluid is taken as ($Pr=0.02$) in all the simulations. The detailed discussions of the results are presented in Section~\ref{sec:reslt_dis}. 

\subsection{Validation and Grid Independence Test \label{sec:valgrid}}
A series of validation studies were conducted to establish the accuracy and robustness of the developed MHD solver. These studies focused on three distinct flow scenarios: (i) Mixed convection without a magnetic field, (ii) Forced convection with an applied magnetic field, and (iii) Buoyancy-driven flow with an applied magnetic field. The discussion on them is presented next.  

Mixed convection without magnetic field: We compared our results to the findings of \citet{sharma2012mixed}, who studied the mixed convection heat transfer over a square cylinder. Table~\ref{tab_validation} presents the mean Nusselt number ($Nu$) for the square cylinder obtained from our study and from \citet{sharma2012mixed} at different Reynolds numbers ($Re$) and Richardson numbers ($Ri$). As the table shows, the Nusselt number values from both studies agree well.
\begin{table}
	\caption{Comparison of Nusselt number (Nu) between the present study and \citet{sharma2012mixed}}
	\centering
	\label{tab_validation}
	\begin{tabular}{c c c c c c}
		\hline
		S. No. & \begin{tabular}[c]{@{}c@{}}Reynolds \\ Number \\ (Re)\end{tabular} & \begin{tabular}[c]{@{}c@{}}Richardson \\ Number\\ (Ri)\end{tabular} & \begin{tabular}[c]{@{}c@{}}Nusselt \\ Number (Nu)\\  Present\end{tabular} & \begin{tabular}[c]{@{}c@{}}Nusselt \\ Number (Nu)\\  \citep{sharma2012mixed}\end{tabular} & Error (\%) \\ \hline
		1      & 30                                                                 & 1                                                                   & 2.914                                                                     & 2.932                                                                              & 0.62      \\ 
		2      & 40                                                                 & 1                                                                   & 3.288                                                                     & 3.297                                                                              & 0.27      \\ 
		3      & 40                                                                 & 0                                                                   & 2.682                                                                     & 2.613                                                                              & 2.64      \\ \hline
	\end{tabular}
\end{table}

\begin{figure}
	\centering
	%  \begin{subfigure}
	%      \centering
	%       \hspace{0.5cm}{\large ($a$)} \hspace{0.7cm} \\
	% \includegraphics[width=0.4\textwidth]{Fig_2a.png} \\
	%      % \hspace{0.5cm}{\large ($a$)} \hspace{0.7cm} \\
	%    % \caption{}
	%      %\label{fig (a): Ux at x=1}
	%  \end{subfigure} 
	%  \begin{subfigure}
	%      \centering
	%      \hspace{0.5cm}{\large ($b$)} \hspace{4.8cm} {\large ($c$)} \hspace{0.5cm} \\
	% \includegraphics[width=0.4\textwidth]{Fig_2b.png}
	%    % \caption{}
	%      %\label{fig (b): Ux at x=1}
	%  \end{subfigure}
	%  \begin{subfigure}
	%      \centering
	% \includegraphics[width=0.4\textwidth]{Fig_2c.png}\\
	%      % \hspace{0.5cm}{\large ($b$)} \hspace{4.8cm} {\large ($c$)} \hspace{0.5cm} \\
	%   %   \caption{}
	%      %\label{fig (b): Ux at x=1}
	%  \end{subfigure}
	%\\
	% {\large ($a$)} \hspace{.5cm} {\large ($b$)} \hspace{2.5cm} {\large ($c$)}\\
	\includegraphics[width=1\textwidth]{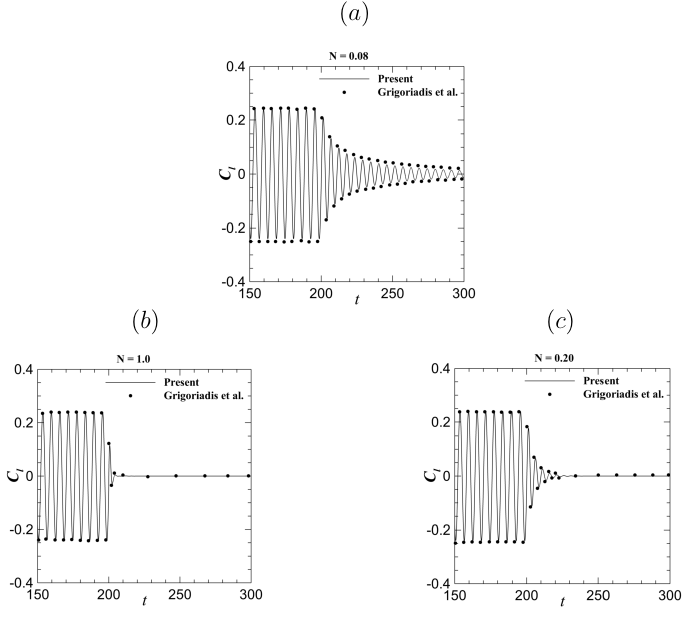}\\
	\caption{Comparison of the time evolution of lift coefficient ($Cl$) at different Stuart number ($N$) between the present study and \citet{grigoriadis2010mhd} is shown in this figure.}
	\label{validation2}        
\end{figure}

Forced convection with an applied magnetic field:  The solver's robustness was further assessed by investigating forced convection flow across a circular cylinder subjected to a magnetic field. The time evolution of the lift force coefficient was compared with the findings of \citet{grigoriadis2010mhd} for various Stuart numbers $(N=Ha^{2}/Re)$. As shown in Figure~\ref{validation2}, a good match is observed between the present results and those of \citet{grigoriadis2010mhd}. Consistent with their approach, the magnetic field was applied after $200$ seconds for each $N$ in our simulations. The comparison of the lift coefficient pre and post-application of the magnetic field with the reference is deemed satisfactory.

\begin{figure}
	\centering
	\includegraphics[width=0.5\textwidth]{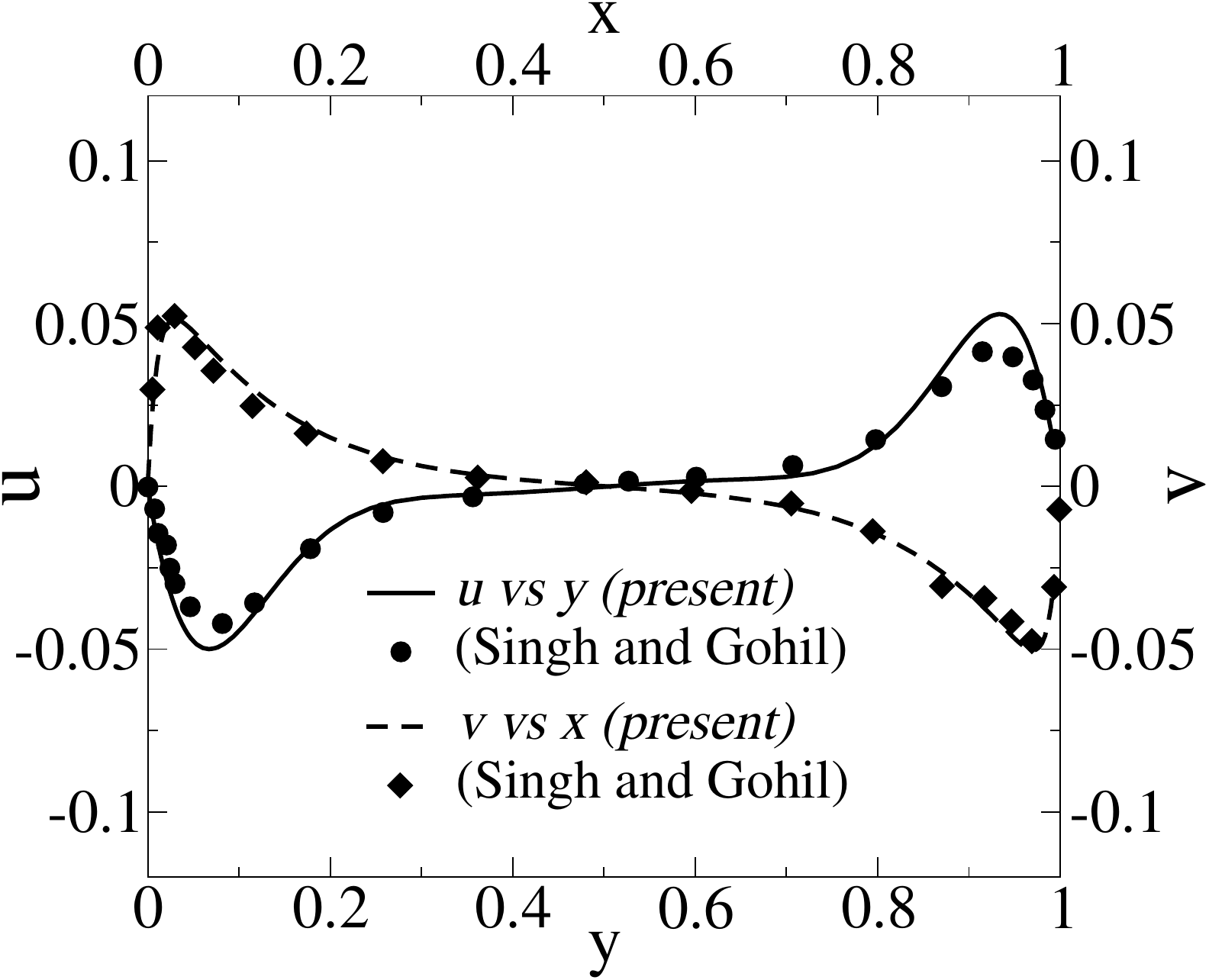}
	\caption{The figure compares the time-averaged velocity components, $u$ and $v$, between the present study and \citet{singh2019numerical}. The comparison is shown at the mid $x-y$ plane for a Rayleigh number ($Ra=10^{6}$) and a Hartmann number ($Ha=100$). The variations of $u$ along the $y$-direction and $v$ along the $x$-direction are presented.}
	\label{validation3}
\end{figure}

Buoyancy-driven flow with an applied magnetic field:  Finally, the solver's capability for simulating natural convection was validated using a cubic cavity exposed to a magnetic field. The computational domain and boundary conditions employed in this validation case were identical to those used by \citet{singh2019numerical}. Figure~\ref{validation3} illustrates the comparison between the horizontal and vertical velocity distributions at the cavity's mid-plane obtained in this study and those reported by \citet{singh2019numerical}. The results demonstrate good agreement between the two sets of data.

These validations demonstrate that the MHD solver produces reliable and accurate results under different flow configurations and conditions. This establishes that the solver is suitable for carrying out the present study of MHD flow past a triangular obstacle with heat transfer (mixed convection). 

\begin{figure}
	\centering
	% \begin{subfigure}
	%     \centering
	%     \hspace{0.7cm}{\large ($a$)} \hspace{4cm} {\large ($b$)}\\
	%     \includegraphics[width=0.4\textwidth]{Fig_4a.eps}
	%     %\caption{$N=0.08$}
	%    % \label{fig (a): Ux at x=1}
	% \end{subfigure} 
	% \begin{subfigure}
	%     \centering
	%     \includegraphics[width=0.4\textwidth]{Fig_4b.eps}
	%    % \caption{$N=0.2$}
	%     %\label{fig (b): Ux at x=1}
	% \end{subfigure}
	% \\
	%   % \hspace{0.7cm}{\large ($a$)} \hspace{5.6cm} {\large ($b$)}\\
	%   \begin{subfigure}
	%     \centering
	%     \hspace{0.7cm}{\large ($c$)} \hspace{4cm} {\large ($d$)}\\
	%     \includegraphics[width=0.4\textwidth]{Fig_4c.eps}
	%     %\caption{$N=0.08$}
	%    % \label{fig (c): Uy at x=1}
	% \end{subfigure} 
	% \begin{subfigure}
	%     \centering
	%     \includegraphics[width=0.4\textwidth]{Fig_4d.eps}
	%    % \caption{$N=0.2$}
	% \end{subfigure}
	\includegraphics[width=1\textwidth]{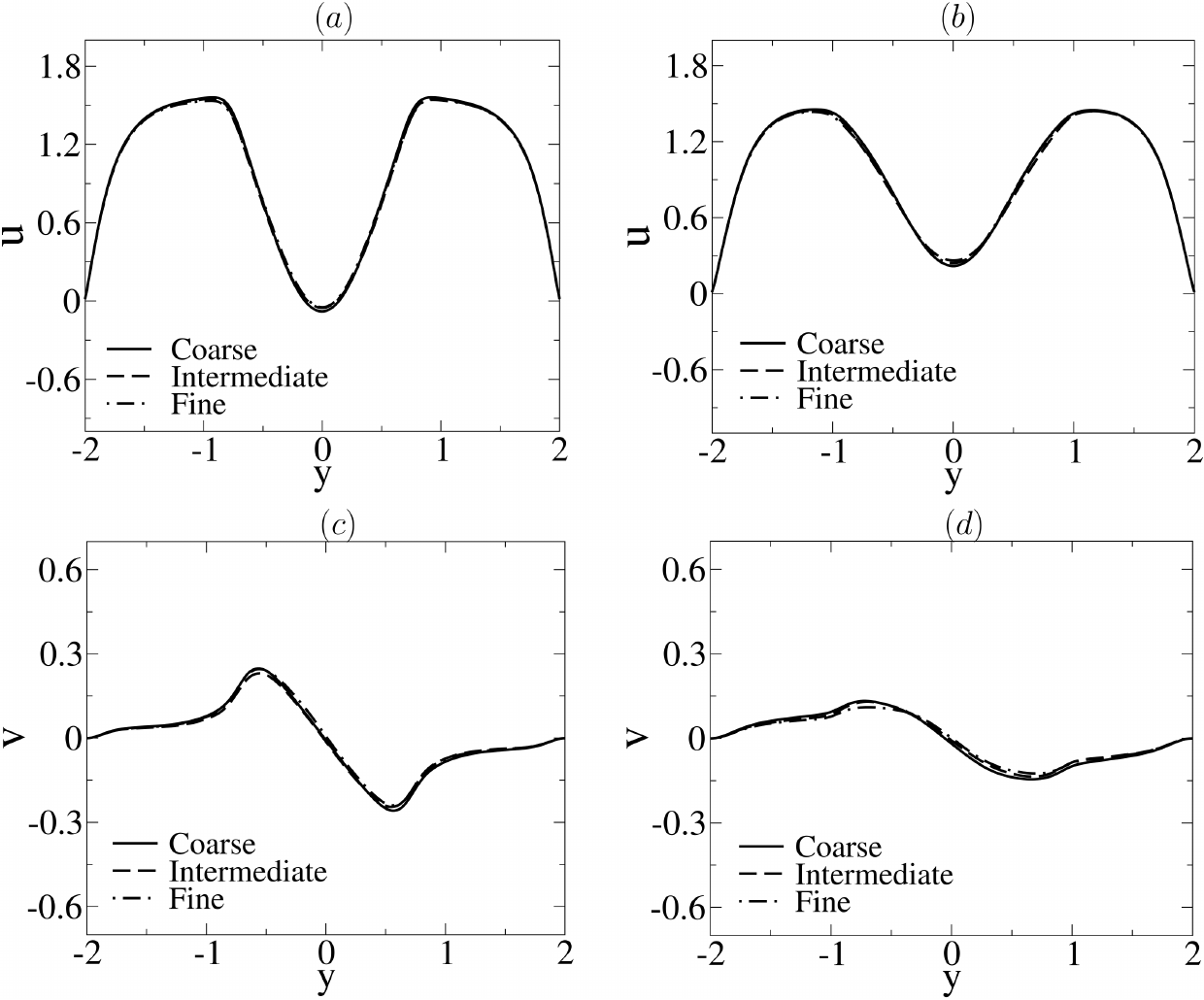}
	% \hspace{0.7cm}{\large ($c$)} \hspace{5cm} {\large ($d$)}\\
	\caption{The figure examines the grid independence test of the simulation on the triangular cylinder with Orientation 1. It shows the variation of the time-averaged velocity components, $u$ (streamwise) and $v$ (transverse), along the $y$-direction at two streamwise positions: $x = 0.25h$ (a \& c) and $x = 0.375h$ (b \& d). The results for three grid sizes are presented: coarse ($1.5$ million cells), intermediate ($2.7$ million cells), and fine ($3.5$ million cells).}
	\label{grid_ind}        
\end{figure}

Similar to the validation study, a grid independence study is conducted to ensure the results (wake dynamics behind the cylinder) are independent of the mesh size. Three grid sizes with varying resolutions were tested: coarse ($1.5$ million cells), intermediate ($2.7$ million cells), and fine ($3.5$ million cells). This test was performed for all three triangular cylinder orientations. We analyzed the time-averaged streamwise velocity ($u$) and transverse velocity ($v$) distributions along the $y$-direction at the mid-plane ($z=0$) for different streamwise locations using the first triangular cylinder orientation. The results are presented in Figure~\ref{grid_ind}. The results from the intermediate and fine grids exhibited good agreement. Therefore, the grid size of $2.7$ million cells (intermediate) was chosen as grid-independent and used for all simulations in this study. Similar analyses are conducted for the other two orientations but not shown for brevity.

\begin{table}
	\centering
	\caption{List of all the cases and their respective parameters for various $Ha$ and $Ri$ values considered within this study at $Re_{h}=600$, with each case examined across all the three orientations: Orientation 1, Orientation 2, and Orientation 3}
	\begin{tabular}{c c c c c}
		\hline
		\textbf{Case} & \textbf{B} & \bm{$Ha$} &\textbf{Gr} & \bm{$Ri$}   \\
		&  $(Tesla)$ &($Bh\sqrt{\frac{\sigma}{\rho\nu}}$) & $(\frac{\beta g \Delta T h^3}{\nu^2})$ & ($\frac{Gr}{Re^2}$)   \\
		\hline
		1 & 0 & 0&  0 & 0\\
		2 & 0 & 0 & $36 \times 10^4$ &1 \\
		3 & 0 & 0 &$18 \times 10^5$ &5 \\
		4 & 0.51031 & 25 & 0 & 0\\
		5 & 0.51031 & 25 & $36 \times 10^4$ &1 \\
		6 & 0.51031 & 25 & $18 \times 10^5$ &5 \\
		7 & 1.02062 & 50 & 0 &0 \\
		8 & 1.02062 & 50 & $36 \times 10^4$ & 1\\
		9 & 1.02062 & 50 & $18 \times 10^5$ &5 \\
		
		\hline
	\end{tabular}
	\label{table1}
\end{table}

\section{Results and Discussion \label{sec:reslt_dis}}
Using the validated numerical model developed, simulations of flow over a triangular prism are carried out for three different orientations. The study investigated the flow and heat transfer characteristics for various $Ha$ and $Ri$ as detailed in Table \ref{table1} while maintaining a constant Reynolds number ($Re_h$) of $600$.
The study begins by discussing the development and evolution of flow for each orientation in the absence of both magnetic field ($Ha=0$) and buoyancy effect ($Ri=0$), i.e., of the baseline case (Case 1). Subsequently, the influence of $Ha$ and $Ri$ on flow characteristics (drag, lift, recirculation length) and heat transfer characteristics (Nusselt number and heat dissipation) are systematically investigated. The behavior of flow three-dimensionality is thoroughly examined and classified, and the point of transition between flow regimes is identified. Finally, the results from all three orientations are compared, and the optimal case with the highest heat transfer is reported.

\subsection{Evolution of flow} \label{evolution of flow}
\begin{figure}
	\centering
	\includegraphics[width=1\textwidth]{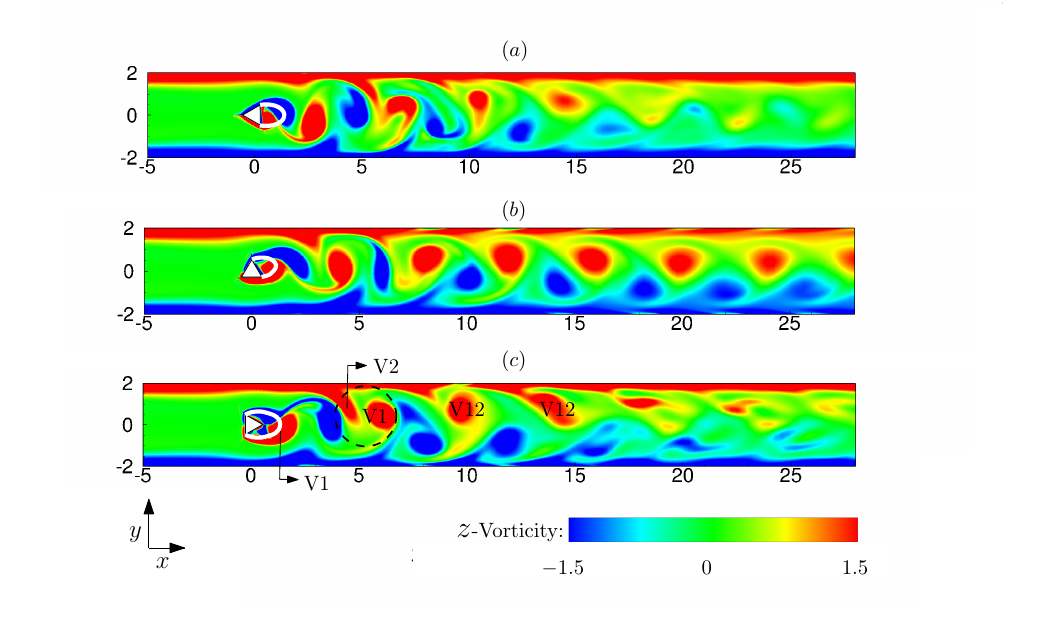}
	%\caption{$z$ Vorticity plots of the flow for $Ha=0$ and $Ri=0$ (case 1) amd the recirculation region is shown by a white line: (a) orientation 1 at $t=350$ (b) orientation 2 at steady state (c) orientation 3 at $t=250$ }
	\caption{$z$ Vorticity plots for all the orientations at baseline case ($Ha=0$, $Ri=0$) with the mean recirculation region indicated by a white line: (a) orientation 1 (b) orientation 2  (c) orientation 3}
	\label{z_vort_c1}
\end{figure}

This section focuses on the flow field evolution for each prism orientation under baseline conditions ($Ha = 0$, $Ri = 0$). As illustrated in Figure~\ref{orientations}, the blockage ratio is $0.25$ for Orientations $1$ and $3$. Orientation $2$ exhibits a slightly lower blockage ratio due to its height being less than the side length. For all three cylinder orientations, the chosen Reynolds number ($Re_{h} = 600$) leads to an unsteady vortex-shedding in the wake of the cylinders. The vortex shedding behavior behind the cylinders is shown in Fig.~\ref{z_vort_c1}.

In contrast to unbounded flows or flows with minimal blockage, where a classic alternate vortex shedding pattern is observed in the wake of a bluff body, the present study with high blockage exhibits distinct shedding behaviors. In the near wake region, an alternating shedding pattern persists. However, further downstream, a transition to a parallel vortex street is observed. This phenomenon of parallel vortex street is consistently observed for all three triangular cylinder orientations considered in the study (Fig.~\ref{z_vort_c1}). This transition to a parallel vortex street is attributed to the gradual interaction between the vortices generated at the channel walls and those shed from the cylinder walls. The channel walls generate vortices of opposite signs compared to the adjacent von K\'{a}rm\'{a}n vortices in the main flow. For illustration, the primary vortex shedding from the bottom side of the cylinder is marked as \textbf{V1}, and the opposite sign vortex originating from the opposite wall of the channel (top wall of the channel) is marked as \textbf{V2}. The opposite-sign vortex, \textbf{V2}, intrudes into the main flow and interacts with the vortex \textbf{V1}, leading to the formation of the vortex \textbf{V12}. As the vortex \textbf{V12} convects downstream, it continues to interact with the vortices from the top channel wall, resulting in a continuous transformation of its shape and strength. A similar process occurs simultaneously for the vortex shedding from the top side of the cylinder and the vortex originating from the bottom channel wall. This interplay between the channel wall vortices and the cylinder wake vortices gradually weakens the alternating shedding pattern, ultimately leading to the observed parallel vortex street behavior further downstream, where opposite-sign vortices are arranged on either side of the centerline.

The observed transition to parallel shedding in the present study aligns with findings from previous experimental and numerical investigations. \citet{griffith2011vortex}, in their three-dimensional study of flow over a circular cylinder at $Re = 300$ within a channel, observed a similar wake pattern. However, they described it as an "\textit{inversion}" of the von K\'{a}rm\'{a}n vortex street. Their work suggests that under specific conditions of blockage ratio and Reynolds number, shed vortices from the cylinder can cross to the opposite side of the channel from where it was originally shedding, leading to the inversion of the vortex street with opposite-sign vortices on either side of the centerline. 
\begin{figure}
	\centering
	\includegraphics[width=1\textwidth]{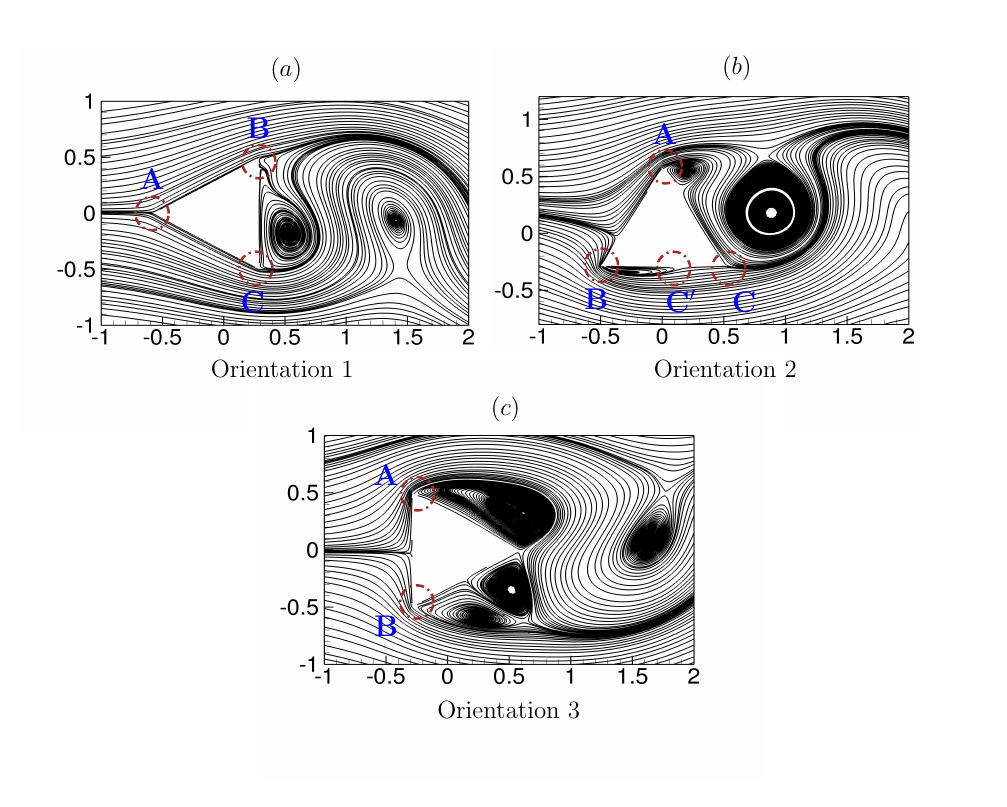}
	\caption{Streamline plots for baseline case ($Ha=0$, $Ri$=0) : $(a)$ Orientation 1 , $(b)$ Orientation 2, $(c)$ Orientation 3.}
	\label{streamlines}
\end{figure}
While all three triangular cylinder orientations ultimately transition to parallel vortex shedding downstream, their near-wake behaviors exhibit distinct characteristics. For Orientation 1, the incoming flow directly impinges the prism and separates symmetrically at the sharp leading edge A (Figure~\ref{streamlines} (a)). The fluid then travels along the inclined walls before separating at the trailing edges B and C, ultimately leading to vortex shedding. In contrast, Orientation 2 presents an asymmetrical obstacle relative to the flow. This asymmetry leads to flow separation initiating at different points (A and B) on the top and bottom surfaces (Figure~\ref{streamlines} (b)). The flow reattaches after separation at point C$'$ and continues along the wall before finally separating at the trailing edge C, resulting in vortex shedding. Orientation 3, similar to Orientation 1, experiences a symmetrical flow separation at the leading edges A and B (Figure~\ref{streamlines} (c)), followed by vortex shedding. Notably, the shed vortices in Orientation 2 display a more regular nature (Fig.~\ref{z_vort_c1}) compared to the other two orientations. This difference can be attributed to the transition to three-dimensional shedding in Orientations 1 and 3, while Orientation 2 retains a two-dimensional shedding behavior.

\begin{figure}
	\centering
	\includegraphics[width=1\textwidth]{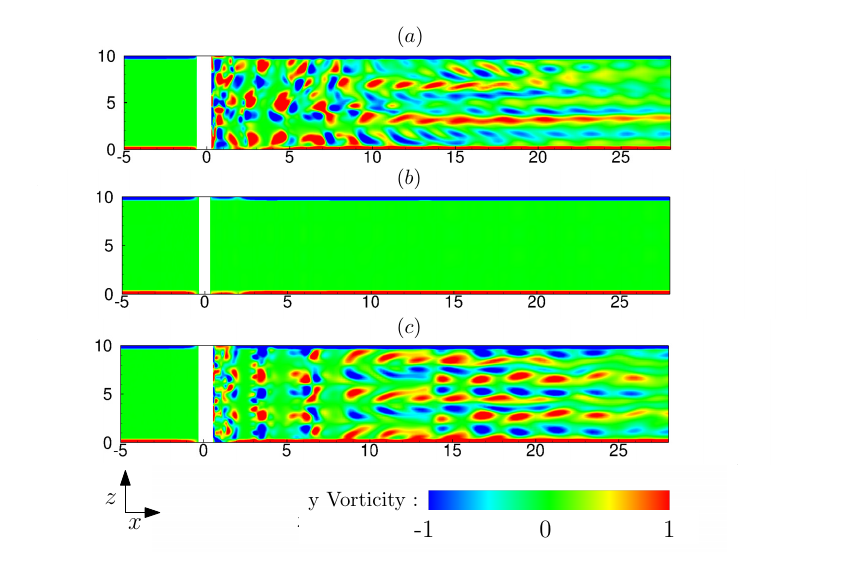}
	\caption{$y$ Vorticity plots for all the orientations at baseline case ($Ha=0$, $Ri=0$): (a) orientation 1 (b) orientation 2  (c) orientation 3}
	\label{y_vort_c1}
\end{figure}

The presence of three-dimensional flow behavior is evident from the y-vorticity plots in Figure~\ref{y_vort_c1}. In contrast to the primary spanwise vortices observed in the $z$-vorticity plots (Figure~\ref{z_vort_c1}), the presence of transverse vorticity ($y$-vorticity) for Orientations 1 and 3 (Figure~\ref{y_vort_c1} ($a$) and ($c$)) confirms the three-dimensionality of the flow in these cases. Instabilities associated with three-dimensional flow behind bluff bodies have been documented in previous studies \citep{zhang1995transition, williamson1996three, saha2003three, rehimi2008experimental}. Depending on their characteristic transverse vorticity wavelengths, these instabilities are classified into Modes A, B, and C. The reported wavelength ranges for these modes are $3-4$ times the characteristic length ($l$) for Mode A, $0.8l-1.5l$ for Mode B, and around $2l$ for Mode C. While Modes A and C can persist throughout the flow, Mode B is typically confined to the near-wake region.

In the present study, the observed periodicity in the transverse direction for Orientations 1 and 3 (Figure~\ref{y_vort_c1} ($a$) and ($c$)) has a wavelength of approximately $4d$, suggesting the presence of a dominant Mode A instability. In unconfined flows over circular and square cylinders, previous research \citep{williamson1996three, saha2003three} has documented the onset of Mode A instabilities at Reynolds numbers ($Re_{d}$, where $d$ is the cylinder diameter/width) in the range of $180-200$. For the present study involving an equilateral triangular cylinder, the corresponding Reynolds number for the three-dimensional transition, based on the cylinder side length ($d$), is $Re_{d} = Ud/\nu = 150$ ($d = h/4$). This earlier transition to three-dimensional flow, compared to past studies, may be attributed to the combined effects of cylinder orientation and the higher blockage ratio employed in this investigation. \citet{rehimi2008experimental} also reported a similar observation of three-dimensional shedding at lower Reynolds numbers for confined flows. The clear evidence of three-dimensionality in Orientations 1 and 3, contrasted with the predominantly two-dimensional behavior observed in Orientation 2 under identical conditions, suggests that Orientation 2 possesses a higher critical Reynolds number for the transition to three-dimensional flow.
\begin{figure}
	\centering
	\includegraphics[width=0.9\textwidth]{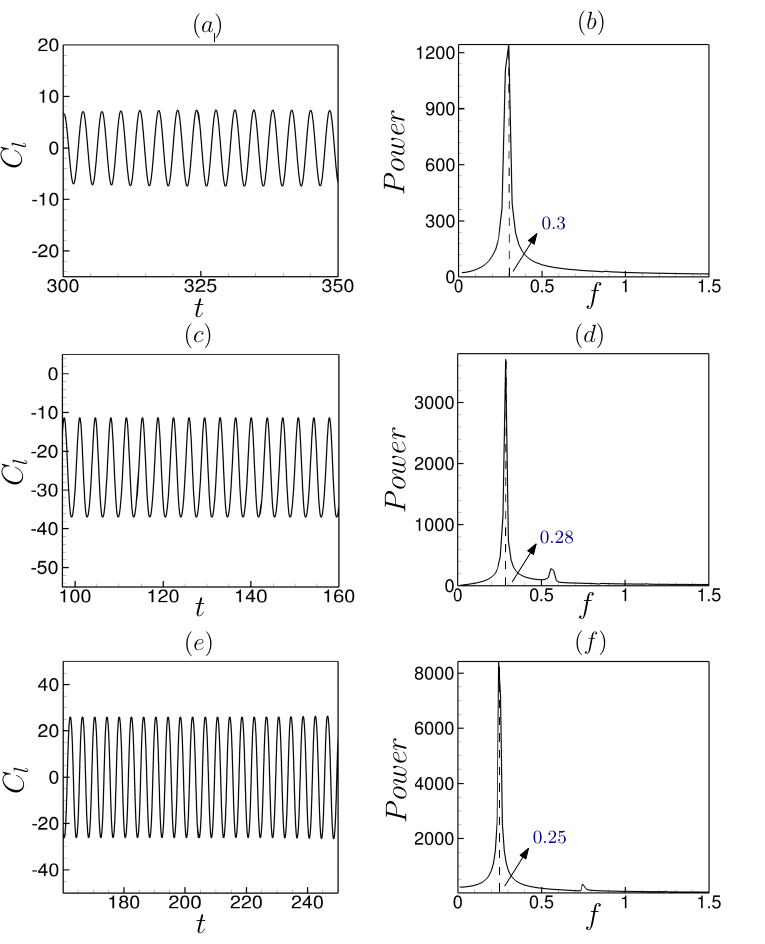}
	\caption{Time history of lift coefficient $C_l$ and its corresponding dominating frequencies for all the orientations at baseline case ($Ha=0, Ri=0$): $(a)$ and $(b)$ correspond to orientation 1, $(c)$ and $(d)$ to orientation 2 and $(e)$ and $(f)$ to orientation 3}
	\label{cl_ft_all_ori}
\end{figure}

While both Orientations 1 and 3 exhibit the transition to three dimensional shedding, closer inspection reveals distinct differences in their wake characteristics (Figs.~\ref{z_vort_c1} and \ref{y_vort_c1}). Notably, Orientation 3 displays a significantly broader recirculation region compared to Orientations 1 and 2. This extended recirculation zone likely contributes to the observed minimum vortex shedding frequency among the three orientations (Figure~\ref{cl_ft_all_ori}). For all three orientations of the cylinder, the alternate shedding of vortices in the wake induces unsteady fluctuations in the pressure field around the cylinder. These temporal variations in the pressure field, acting on the cylinder surfaces in both the streamwise ($x$) and transverse ($y$) directions, result in corresponding fluctuations in the drag and lift coefficients over time. Figures~\ref{cl_ft_all_ori}(a), (c), and (e) illustrate the time history of the lift coefficient ($C_l$) for the different orientations. The vortex shedding frequency is calculated using Fast Fourier Transformation (FFT) applied to the time-series data of the lift force acting on the cylinder (Figure~\ref{cl_ft_all_ori}). The analysis reveals shedding frequencies of $0.3$, $0.28$, and $0.25$ for Orientations 1, 2, and 3, respectively. The lower shedding frequency observed in Orientation 3 can be attributed to the formation of larger vortices from the cylinder surfaces. These larger vortices require a longer time for formation and subsequent separation from the cylinder surface, ultimately leading to a lower shedding frequency. These observations are consistent with the findings of \citet{nakagawa1989vortex} on high Reynolds number ($Re = 1.73 \times 10^{5}$) flow over a triangular prism. Their study, which investigated two orientations of a triangular prism (corresponding to Orientations 1 and 3 in the present study), also reported a lower vortex shedding frequency for Orientation 3 compared to Orientation 1. 

\begin{figure}
	\centering
	\includegraphics[width=1\textwidth]{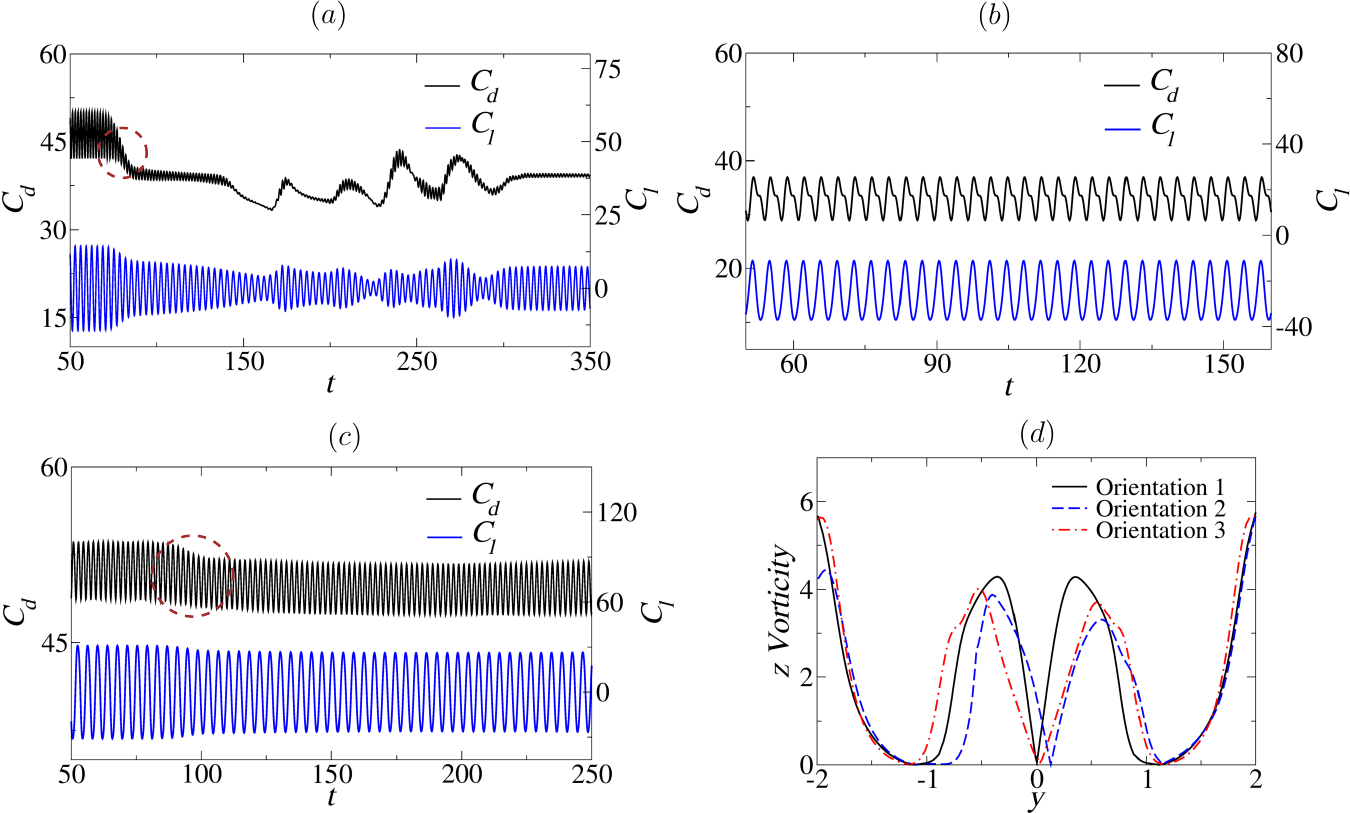}
	%\caption{Variation of $C_d$, $C_l$ and $z$ mean vorticity for baseline case of all the three orientations $(a)$ Orientation 1; $(b)$ Orientation 2; $(c)$ Orientation 3; $(d)$ $z$ mean vorticity taken along $y$ direction at a distance of $0.25h$ from the prism centroid}
	\caption{The figures $(a)$, $(b)$ and $(c)$ illustrates the time history of $C_d$ and $C_l$  at the beseline case ($Ha=0$, $Ri=0$) for : ($a$) orientation 1, ($b$) orientation 2 and $(c)$ orientation 3. Figure $(d)$ shows the variation of mean $z$ vorticity taken along $y$ direction at a distance of $0.25h$ from the prisms centroid}
	\label{cdcl_c1}
\end{figure}

The drag coefficient ($C_d$) and lift coefficient ($C_l$) for all three orientations are presented in Figure~\ref{cdcl_c1}. While Orientations 1 and 3 exhibit a net lift force of zero on the cylinder, Orientation 2 experiences a sustained downward net lift force indicating the asymmetric nature of flow for this configuration. This asymmetry in the flow behavior of Orientation 2 can be attributed to its asymmetric orientation relative to the incoming flow. This configuration leads to the shedding of two unequal-sized vortices from either side of the cylinder surface. This presence of unequal-sized vortices of different strengths on either side of the centerline for Orientation 2 is confirmed from the mean $z$-vorticity plot in Fig.~\ref{cdcl_c1} (d). As these vortices shed, they induce an unequal entrainment of fluid from either side of the cylinder in the transverse direction. This unequal entrainment ultimately results in the observed asymmetric flow behavior and a net non-zero lift force in Orientation 2.

The analysis of the drag coefficient ($C_d$) plots reveals the highest mean value for Orientation 3 and the lowest for Orientation 2 (Figure~\ref{cdcl_c1}). The larger drag observed in Orientation 3 can be attributed to the presence of a wider and more extensive recirculation region, which aligns with the findings of \cite{nakagawa1989vortex}.  Additionally, the study by \citet{agrwal2016experimental} suggests that the formation of larger vortices leads to a wider recirculation zone and increased wake width, further supporting the observations in the present study. While Orientation 3 exhibits the largest wake width, followed by Orientation 1 and then Orientation 2, the drag coefficient is lowest for Orientation 2. This observation may be attributed to the two-dimensional nature of the flow in Orientation 2. The time-series data of $C_d$ for Orientations 1 and 3 also clearly illustrates the transition from two-dimensional to three-dimensional flow over time. This transition is marked by a noticeable drop in the $C_d$ value and is highlighted by red dotted circles in Figure~\ref{cdcl_c1} (a) and (c).

\subsection{Influence of $Ha$ and $Ri$ on flow pattern}
In this section, we explore the effects of the applied magnetic field strength ($Ha$) and the buoyancy force parameter ($Ri$) on the wake behavior of the triangular cylinders across all three orientations. An increase in $Ha$ leads to a corresponding rise in the Lorentz force, which acts in opposition to the inertial forces within the flow. This counteracting effect can be observed as a damping phenomenon on vortices oriented at an angle relative to the applied magnetic field. Consequently, the wake characteristics may deviate from those observed in the absence of a magnetic field. Similarly, an increase in $Ri$ signifies a greater influence of buoyancy forces within the flow. These buoyancy forces can promote mixing within the fluid, leading to potential changes in the flow patterns observed in the wake region behind the cylinders. The specific nature of these changes will depend on the interplay between buoyancy forces, inertial forces, applied magnetic field strength, and cylinder orientation.

\begin{figure}
	\centering
	\includegraphics[width=1\textwidth]{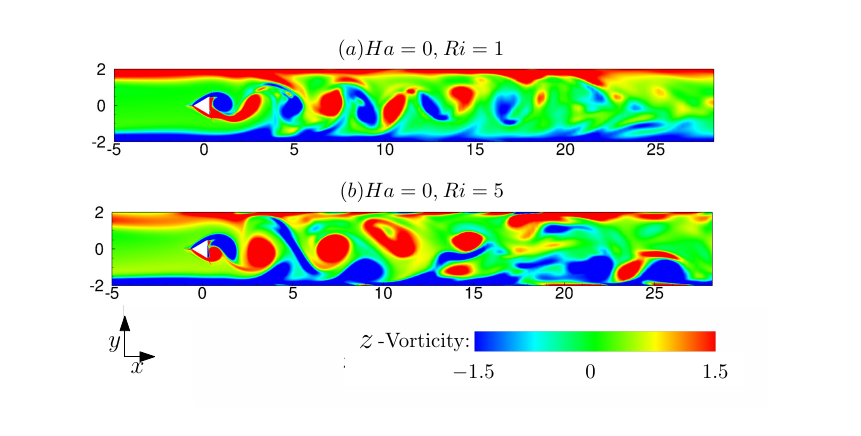}
	\caption{$z$ Vorticity plots for orientation 1 at $Ha=0$ with varying $Ri$: $(a)$ $Ri=1$, $(b)$ $Ri=5$}
	\label{z_vort_ha0m1_ri1ri5}
\end{figure}

\begin{figure}
	\centering
	\includegraphics[width=1\textwidth]{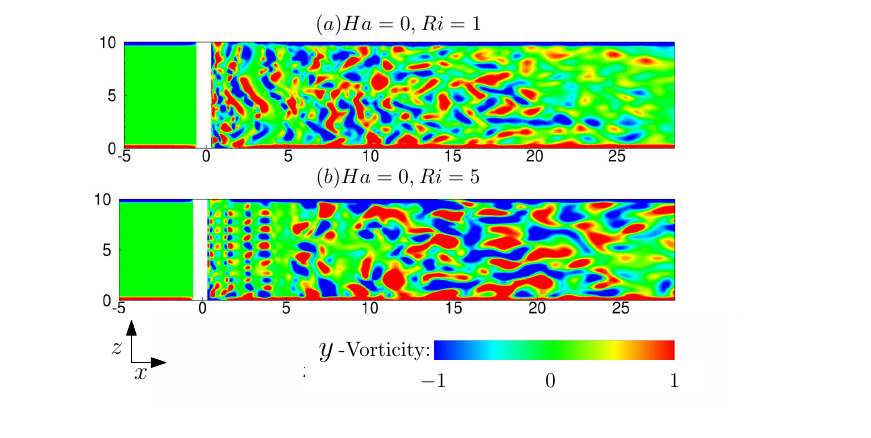}
	\caption{$y$ Vorticity plots for orientation 1 at $Ha=0$ with varying $Ri$: $(a)$ $Ri=1$, $(b)$ $Ri=5$}
	\label{y_vort_ha0m1_ri1_5}
\end{figure}

\begin{figure}[h!]
	\centering
	\includegraphics[width=1\textwidth]{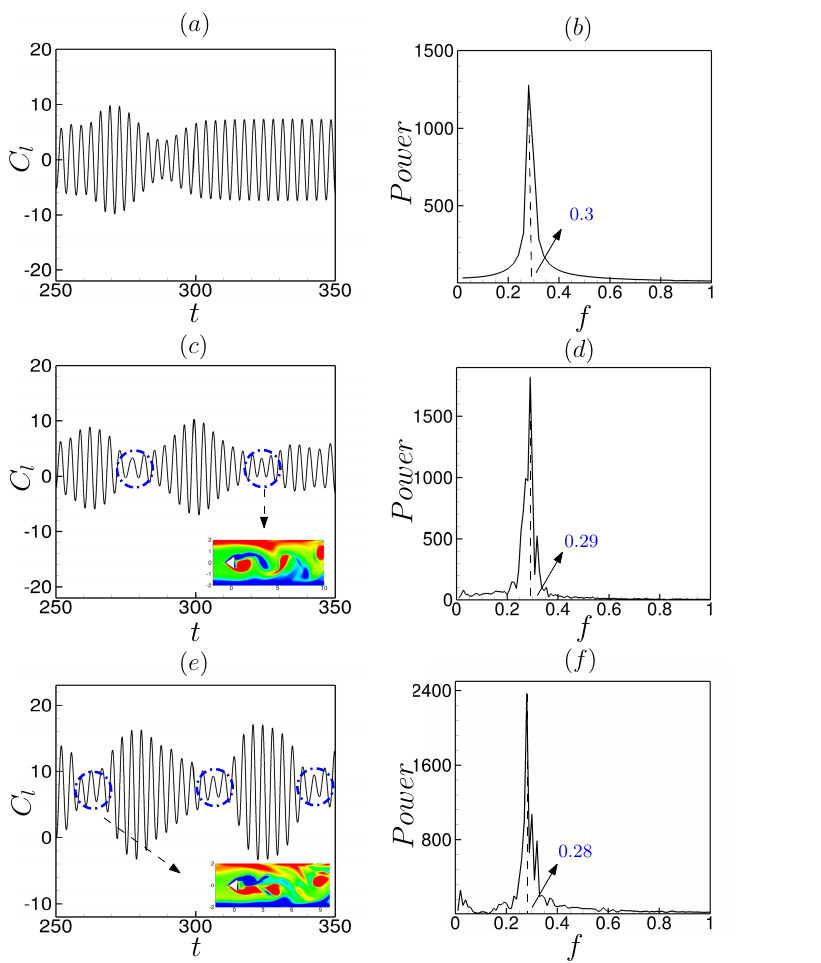}
	\caption{Time history of lift coefficient $C_l$ and its corresponding frequencies for orientation 1 with insets of $z$ vorticity contours for the corresponding case at $Ha=0$ with varying $Ri$ : $(a)$ and $(b)$ correspond to $Ri=0$, $(c)$ and $(d)$ to $Ri=1$ and $(e)$ and $(f)$ to $Ri=5$}
	\label{cl_ft_ha0m1}
\end{figure}

For Orientation 1 at zero magnetic fields ($Ha = 0$) (refer to cases 1, 2, and 3 in Table~\ref{table1}), increasing buoyancy forces ($Ri$) significantly impact the wake behavior. As $Ri$ increases, buoyancy effects become more prominent near the heated bottom wall of the channel, leading to the formation of stronger wall vortices in this region (Figure~\ref{z_vort_ha0m1_ri1ri5}). Here onward, these channel-wall vortices will be referred to as "wall vortices" for brevity. The interaction between these strengthened wall vortices and the shed vortices in the wake plays a significant role in altering the wake characteristics. This interaction enhances the wake vortices adjacent to the bottom wall, and the continuous interplay further downstream leads to the development of a complex, chaotic flow regime (Figure~\ref{z_vort_ha0m1_ri1ri5}). While all three cases exhibit three-dimensional wake behavior, the specific instability modes associated with these flows differ.

The $y$-vorticity plot for Case 2 ($Ha = 0$, $Ri = 1$) (Figure~\ref{y_vort_ha0m1_ri1_5} (a)) exhibits a "chaotic regime" with wavelength ranging from approximately $1.5d$ to $2d$. Unlike the well-defined and periodic structures associated with pure Mode A or B instabilities, the "chaotic region" in Case 2 displays a more irregular and mixed pattern of alternating positive and negative vorticity zones. This observation aligns with the findings of \citet{zhang1995transition} who reported a similar chaotic region during the transition between Mode A and Mode B instabilities in their study of flow over a circular cylinder. The presence of this chaotic region in Case 2 signifies the influence of the enhanced wall vortices (due to increased $Ri$) on the three-dimensional flow behavior. The interaction between the wall vortices and the inherent instabilities within the flow likely led to this complex transitional state between Mode A and Mode B. In contrast to Case 2, Case 3 ($Ha = 0$, $Ri = 5$) exhibits a dominant Mode B instability with a wavelength of $2d$ (Figure~\ref{y_vort_ha0m1_ri1_5} (b)). Furthermore, the Mode B instability is confined to the near-wake region and aligns with established knowledge of this mode.

Beyond the observed variations in instability modes associated with the three-dimensional wake, the flow exhibits an additional instability that leads to the intermittent delay of vortex shedding behind the cylinder for cases at $Ri=1$ and $Ri=5$. This phenomenon is confirmed by the time-series plots of the lift coefficient ($C_l$) in Figure~\ref{cl_ft_ha0m1}. Here, intermittent periods of low-amplitude fluctuations in $C_l$ correspond to instances of delay in vortex shedding. The insets in Figure~\ref{cl_ft_ha0m1} (c, e) showcase $z$-vorticity contours corresponding to the low $C_l$ regions observed in the time series. These contours suggest a potential mechanism for the observed low $C_l$ fluctuations. 
% \Amulya{At $Ri=0$, the low-amplitude fluctuations and the corresponding contour indicate the absence of intermittent delay in vortex shedding and are instead associated with a reduction in the strength of the shedding vortices. Whereas, at $Ri=1$ and $Ri=5$,}  
During these low $C_l$ periods, the counter-rotating vortices forming behind the cylinder exhibit an elongated morphology compared to those observed during regular shedding (Figure~\ref{cl_ft_ha0m1}). This elongation likely delays the detachment process of the vortices, leading to a temporary hindrance in the vortex shedding frequency and the observed low amplitude fluctuations in the lift coefficient. The plots in Figure~\ref{cl_ft_ha0m1} also reveal a trend of the frequent occurrence of these low-amplitude $C_l$ fluctuations with increasing $Ri$. This suggests that the delay of vortex-shedding events becomes more frequent as buoyancy forces become stronger. Furthermore, the results of the FFT applied to the $C_l$ time-series data show a decrease in the dominant vortex shedding frequency as $Ri$ increases (Figures~\ref{cl_ft_ha0m1} (b, d, f)). Additionally, the presence of mixed frequencies in the FFT spectra for $Ri = 1$ and $5$ (Figures~\ref{cl_ft_ha0m1} (d and f)) aligns with the earlier observation of a complex and chaotic flow regime at higher $Ri$ values. This reinforces the notion that increasing buoyancy forces significantly influences the stability of the wake and the regularity of vortex shedding behavior.

\begin{figure}[h!]
	\centering
	\includegraphics[width=1\textwidth]{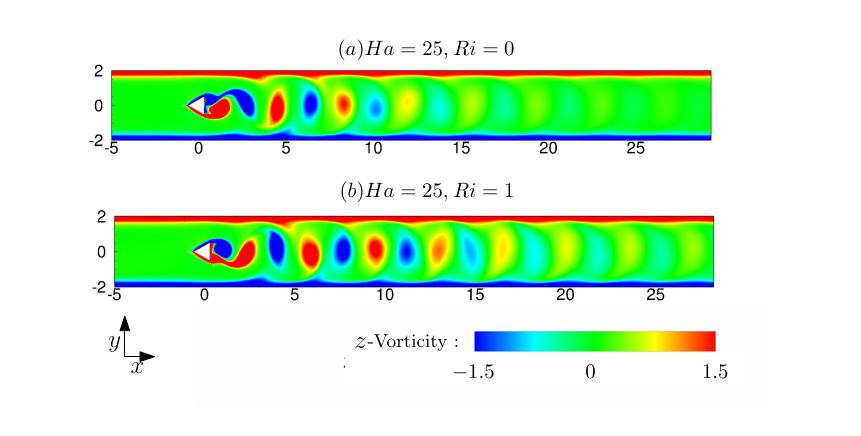}
	\caption{$z$ Vorticity plots for orientation 1 at $Ha=25$ with varying $Ri$: $(a)$ $Ri=0$, $(b)$ $Ri=1$}
	\label{zvort_ri0ri1m1_ha25}
\end{figure}
\begin{figure}[h!]
	\centering
	\includegraphics[width=1\textwidth]{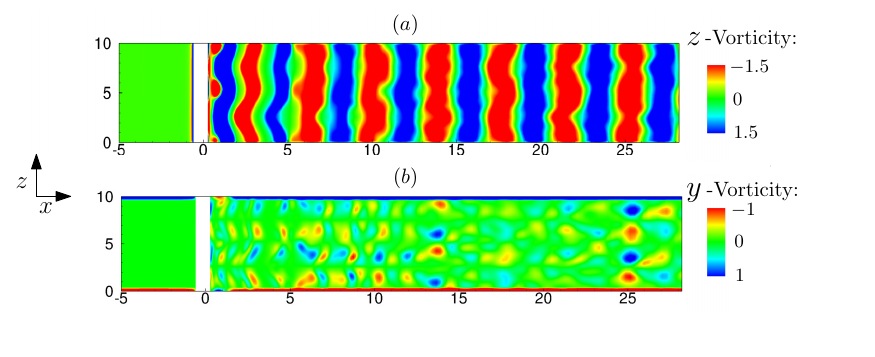}
	\caption{Vorticity plots in $x-z$ plane for orientation 1 at $Ha=25$ and $Ri=5$ (a) $z$ Vorticity and (b) $y$ Vorticity}
	\label{ynz_ha25m1_ri5}
\end{figure}

The influence of the magnetic field strength ($Ha$) on the wake is evident at higher $Ha$ values. For Cases 4 ($Ha=25$, $Ri=0$) and 5 ($Ha=25$, $Ri=1$) (Figures~\ref{zvort_ri0ri1m1_ha25}), the increased Lorentz forces due to the stronger magnetic field suppress both inertial and buoyancy effects. This suppression leads to a transition from the previously observed three-dimensional flow to a two-dimensional vortex-shedding behavior (Figure~\ref{zvort_ri0ri1m1_ha25}). Notably, this resembles the classical two-dimensional von K\'{a}rm\'{a}n vortex street, distinct from the two-dimensional parallel vortex street observed earlier at lower $Ha$ ($Ha=0$). 

However, for Case 6 ($Ha = 25$, $Ri = 5$), the interplay between forces becomes more complex. While the magnetic field remains strong, the high buoyancy ($Ri = 5$) partially counteracts the Lorentz force suppression. This results in a small yet noticeable presence of $y$-vorticity in the near wake region (Figure~\ref{ynz_ha25m1_ri5} (b)), indicating a nascent transition back to three-dimensionality. The $z$-vorticity plot in Figure~\ref{ynz_ha25m1_ri5} (a) reveals that the vortices appear to adhere to the cylinder surface, a phenomenon classified as the "adhesion mode" by \citet{zhang1995transition}. They reported this mode to occur around critical Reynolds numbers, with regions between adhesion points exhibiting Mode A patterns. This aligns with our observations, where the transverse vorticity wavelength with a value of approximately $4d$ suggests a Mode A pattern. Finally, at an even stronger magnetic field strength ($Ha = 50$), all cases ($7, 8,$ and $9$) exhibit complete suppression of vortex shedding (figures not shown). Here, the dominant influence of the magnetic field overwhelms both inertial and buoyancy forces, leading to minimal to no observable difference in flow dynamics with varying $Ri$ values.

\begin{figure}[h]
	\centering
	\includegraphics[width=1\textwidth]{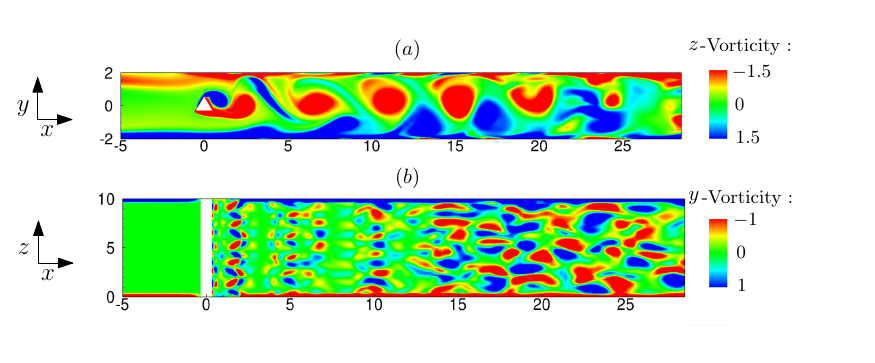}
	\caption{Vorticity plots for orientation 2 at $Ha=0$ and $Ri=5$ (a) $z$ Vorticity in $x-y$ plane and (b) $y$ Vorticity in $x-z$ plane}
	\label{ynz_ha0ri5m2}
\end{figure}
\begin{figure}[h]
	\centering
	\includegraphics[width=1\textwidth]{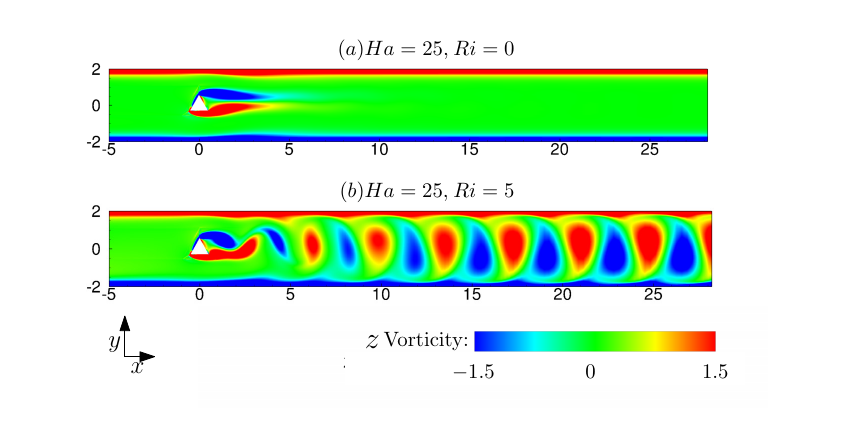}
	\caption{$z$ Vorticity plots for orientation 2 at $Ha=25$ with varying $Ri$: $(a)$ $Ri=0$, $(b)$ $Ri=5$}
	\label{z_vort_m2_ha25_ri0ri5}
\end{figure}
Orientation 2 exhibits distinct flow behavior variations with changes in $Ha$ and $Ri$. In cases with $Ha = 0$, $Ri = 0$ and $1$, a two-dimensional vortex shedding pattern is observed (refer to cases 1 and 2). However, when buoyancy forces become significant ($Ri = 5$), the flow transitions to a three-dimensional wake (case 3). The instability mode associated with this three-dimensional wake displays characteristics of Mode C, with a transverse vorticity wavelength of approximately $2d$ (Figure~\ref{ynz_ha0ri5m2} (b)). Notably, unlike Orientation 1, where intermittent vortex shedding delay is observed at $Ha = 0$, $Ri=1~\&~5$, Orientation 2 does not exhibit this phenomenon under similar conditions.

Increasing $Ha$ to $25$ reveals a more prominent influence of the magnetic field. For $Ri = 0$ (case 4, Figure~\ref{z_vort_m2_ha25_ri0ri5} (a)), the magnetic field suppresses vortex shedding entirely. A weak shedding behavior persists for $Ha = 25$ and $Ri = 1$ (not shown). However, for $Ha = 25$ and $Ri = 5$ (case 6, Figure~\ref{z_vort_m2_ha25_ri0ri5} (b)), the wake transitions to a classical two-dimensional von K\'{a}rm\'{a}n vortex street. This behavior contrasts with the parallel two-dimensional vortex street observed at $Ha = 0$ for all $Ri$ values in Orientation 2. Finally, at the strongest magnetic field ($Ha = 50$), all cases (7, 8, and 9) exhibit complete vortex shedding suppression, with minimal to no observable difference in the flow dynamics across varying $Ri$ values for Orientation 2.

\begin{figure}[h]
	\centering
	\includegraphics[width=1\textwidth]{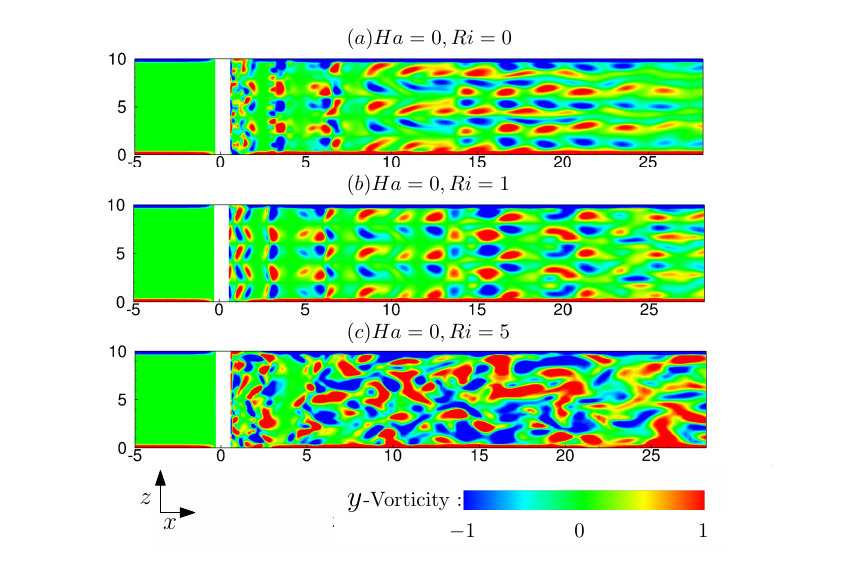}
	\caption{$y$ Vorticity plots for orientation 3 at $Ha=0$ with varying $Ri$: $(a)$ $Ri=0$, $(b)$ $Ri=1$ and $(b)$ $Ri=5$  }
	\label{y_vort_m3}
\end{figure}
Orientation 3 exhibits wake behavior similar to that observed in Orientation 1 with increasing $Ha$ and $Ri$. All cases with $Ha = 0$ (cases 1, 2, and 3) demonstrate three-dimensional flow characteristics (Figure~\ref{y_vort_m3}). Notably, only case 3 ($Ri = 5$) displays a delay in vortex shedding behavior, similar to Orientation 1. Cases 1 and 2 exhibit transverse vorticity wavelengths of approximately $4d$, suggesting dominant Mode A instabilities. In contrast, case 3 displays a wavelength near $2d$ but with flow features resembling a chaotic region, indicative of a potential mixture of Mode A and Mode B instabilities. For $Ha = 25$ (cases 4, 5, and 6), regardless of $Ri$ value, the wake transitions to a two-dimensional state. Finally, at the strongest magnetic field ($Ha = 50$), all cases (7, 8, and 9) exhibit complete vortex shedding suppression, aligning with the observations in Orientations 1 and 2.

\begin{figure}[h]
	\centering
	%      \begin{subfigure}
	%          \centering
	%          \hspace{0.7cm}{\large ($a$)} \hspace{5.2cm} {\large ($b$)}\\
	%          \includegraphics[width=0.4\textwidth]{Fig_18a.eps}
	%          %\caption{$N=0.08$}
	%          % \label{}
	%      \end{subfigure} 
	%      \begin{subfigure}
	%          \centering
	%          \includegraphics[width=0.4\textwidth]{Fig_18b.eps}
	%         % \caption{$N=0.2$}
	%          % \label{cd_vs_ri}
	%      \end{subfigure} \\
	%    % \hspace{0.7cm}{\large ($c$)} \hspace{5.6cm} {\large ($d$)}\\
	%         % \caption{caption}
	%         % \label{grid_ind}        
	% % \end{figure}
	% % \begin{figure}
	% %      \centering
	%      \begin{subfigure}
	%          \centering
	%          \hspace{0.7cm}{\large ($c$)} \hspace{5.2cm} {\large ($d$)}\\
	%          \includegraphics[width=0.4\textwidth]{Fig_18c.eps}
	%          %\caption{$N=0.08$}
	%          % \label{cl_vs_ha}
	%      \end{subfigure} 
	%      \begin{subfigure}
	%          \centering
	%          \includegraphics[width=0.4\textwidth]{Fig_18d.eps}
	%         % \caption{$N=0.2$}
	%          % \label{cl_vs_ri}
	%      \end{subfigure}
	% \hspace{0.7cm}{\large ($c$)} \hspace{5.6cm} {\large ($d$)}\\
	\includegraphics[width=1\textwidth]{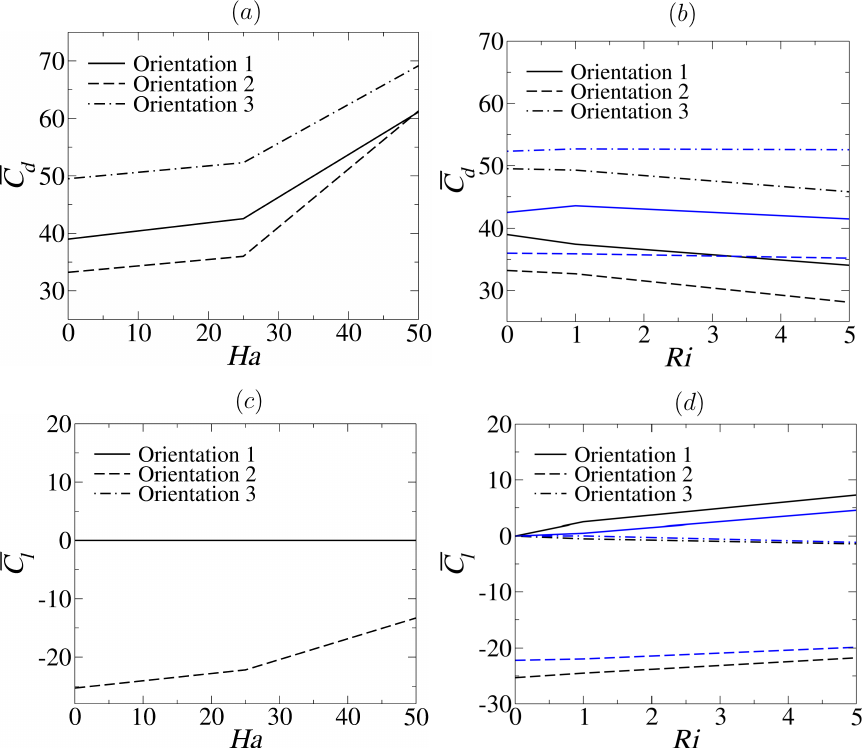}
	\caption{Variation of mean drag and lift coefficient with $Ha$ and $Ri$ for all the three orientations of the triangular prism. Figures (a) and (c) show the variations with $Ha$ at $Ri = 0$, while (b) and (d) show the variations with $Ri$ at $Ha = 0$ (black) and $Ha = 25$ (blue) }
	\label{cdcl_variations}        
\end{figure}
% \begin{figure}
% \centering
% \includegraphics{buoyancy_forces.eps}
% \caption{Action of buoyancy forces ($F_B$) on the prism for all three orientations, while the streamline flow is from left to right}
% \label{buoyancy_forces}
% \end{figure}
Figure~\ref{cdcl_variations} explores the influence of $Ha$ and $Ri$ on the mean drag coefficient ($\overline{C_d}$) and mean lift coefficient ($\overline{C_l}$). $\overline{C_d}$ is observed to increase with increasing $Ha$ at constant $Ri$ for all cylinder orientations. However, the plot for $Ri=0$ with different values of $Ha$ is given in Fig.~\ref{cdcl_variations} (a), where $\overline{C_d}$ increases with an increase in $Ha$. This aligns with findings reported by \cite{muck2000three, chatterjee2014wall, chatterjee2012control,sekhar2007effect} for both three-dimensional and two-dimensional magnetohydrodynamic (MHD) flows over obstacles. Furthermore, a two-dimensional study of transverse MHD flow by \citet{singha2011control} also confirmed that the drag coefficient rises more rapidly at higher Hartmann numbers, which is in agreement with the present observations. Figure~\ref{cdcl_variations} (b) depicts the variation of $\overline{C_d}$ with $Ri$ at constant $Ha$. Here, $\overline{C_d}$ decreases with increasing $Ri$. This can be attributed to the enhanced buoyancy effects at higher $Ri$. The increased buoyancy causes the flow to rise from the heated bottom wall, promoting stronger mixing and consequently increasing the local velocity magnitude. This rise in the local Reynolds number decreases the pressure field, resulting in a reduced drag force acting on the cylinder. Similar observations of decreasing drag coefficient with increasing Reynolds number were reported by \cite{sharma2004heat} for forced convection flows over a square cylinder. As anticipated, the $\overline{C_l}$ for Orientations 1 and 3 exhibits a value of zero at $Ri = 0$ across all Ha values (Figure~\ref{cdcl_variations} (c)). This aligns with the symmetrical nature of their geometries, resulting in no net lift force in the absence of buoyancy effects. Orientation 2, due to its inherent asymmetry, experiences a non-zero mean lift force at $Ha=0$ and $Ri=0$. The magnitude of this force is found to decrease with increasing $Ha$. An increase in magnetic field strength weakens the vortices and controls any asymmetry in the flow, thereby decreasing the mean lift. Interestingly, for Orientation 1, increasing $Ri$ at a constant $Ha$ leads to a net lift force that continues to rise with $Ri$. A similar trend of increasing $\overline{C_l}$ with $Ri$ is observed for Orientation 2. In contrast, Orientation 3 displays a negligible mean lift force with increasing $Ri$.

\begin{figure}[h]
	\centering
	% \begin{subfigure}
	%     \centering
	%      {\hspace{0.6cm}\large ($a$)} \hspace{4cm} {\large ($b$)} \hspace{0.2cm} \\
	%     \includegraphics[width=0.4\textwidth]{Fig_19a.eps} 
	% \end{subfigure} 
	% \begin{subfigure}
	%     \centering
	%     \includegraphics[width=0.4\textwidth]{Fig_19b.eps}\\
	% \end{subfigure}
	% \begin{subfigure}
	%     \centering
	%     {\hspace{0.6cm}\large ($c$)} \hspace{4cm} {\large ($d$)} \hspace{0.2cm} \\
	%     \includegraphics[width=0.4\textwidth]{Fig_19c.eps}
	% \end{subfigure}
	% \begin{subfigure}
	%     \centering
	%     \includegraphics[width=0.4\textwidth]{Fig_19d.eps}\\
	% \end{subfigure}
	%\\
	% {\large ($a$)} \hspace{.5cm} {\large ($b$)} \hspace{2.5cm} {\large ($c$)}\\
	\includegraphics[width=1\textwidth]{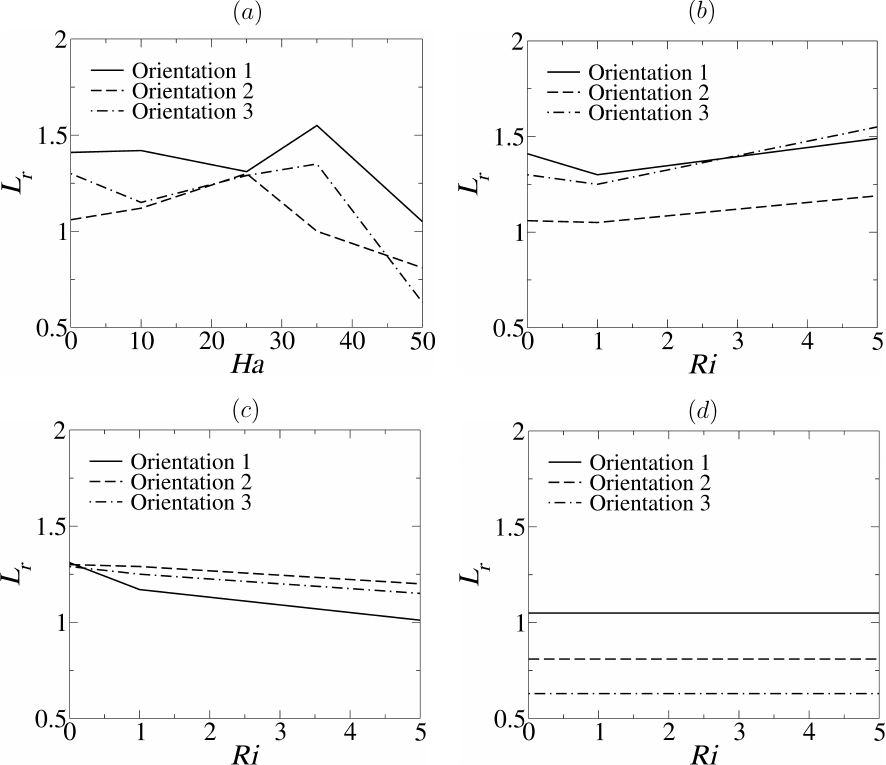}
	\caption{Variation of recirculation length with $Ha$ and $Ri$ for all the three orientations of triangular prism : Figure $(a)$ shows the variation with $Ha$ at $Ri=0$, Figures $(b)$, $(c)$ and $(d)$ show the variation with $Ri$ at $Ha=0$, $Ha=25$ and $Ha=50$ respectively.}  
	\label{lr_plot_4}
\end{figure}

The influence of the magnetic field strength ($Ha$) and buoyancy effects ($Ri$) on the mean recirculation length ($L_r$) is investigated further. $L_r$ is defined as the horizontal distance from the prism's center to the point where the streamwise velocity changes direction. The variation of mean recirculation length with $Ha$ and $Ri$ for all three orientations is shown in Figure~\ref{lr_plot_4}. For Orientations 1 and 3, where the wake behind the cylinders shows three-dimensional behavior at $Ha=0~\&~Ri=0$, the $L_r$ decreases with increasing $Ha$ until the wake transitions to a two-dimensional wake. Further increasing $Ha$ increases the $L_r$ due to the elongation of the shear layers until a critical $Ha$ is reached, where vortex shedding is completely suppressed. Beyond this point, increasing $Ha$ causes $L_r$ to decrease (Figure~\ref{lr_plot_4}(a)). Orientation 2 with initial two-dimensionality at $Ha = 0~\&~Ri = 0$ shows a similar trend, with $L_r$ initially increasing till the critical $Ha$ and then decreasing with a further increase in $Ha$. \citet{chatterjee2013wall} analyzed MHD flow over a circular cylinder and reported a decrease in $L_r$ with increasing $Ha$ in steady flow conditions similar to the present study. 

The variation of mean $L_r$ with $Ri$ exhibits distinct behaviors for different $Ha$ values. At $Ha = 0$ (Figure~\ref{lr_plot_4}(b)), a slight decrease in $L_r$ is followed by a subsequent increase for all orientations with increasing $Ri$. This can be attributed to the increased buoyancy forces that elongate $L_r$ and elevate the wake region. In contrast, at $Ha = 25$ (Figure~\ref{lr_plot_4}(c)), $L_r$ is observed to decrease with increasing $Ri$ for all orientations due to the action of Lorentz forces suppressing the streamwise velocity and causing $L_r$ to decrease. Interestingly, cases with high magnetic field strength ($Ha = 50$) display a constant $L_r$ irrespective of $Ri$ variations (Figure~\ref{lr_plot_4}(d)) due to the complete suppression of vortex shedding.

\subsection{Influence of $Ha$ and $Ri$ on Heat Transfer} \label{heat_transfer}
This section explores the interplay between the applied magnetic field strength ($Ha$), buoyancy force parameter ($Ri$), and cylinder orientation on heat transfer behavior within the flow domain. Bluff bodies, like the cylinders in this study, are known to promote heat transfer through vortex shedding, which enhances the mixing between hot and cold fluids. Similarly, buoyancy forces can lift hot fluids, further promoting their interaction with colder regions and increasing heat transfer. The applied magnetic field introduces an additional factor -- Lorentz forces -- which influence the overall flow behavior and, consequently, heat transfer characteristics. Therefore, the heat transfer observed in this study is a result of the complex interplay between buoyancy forces (influencing vertical movement), inertial forces (related to fluid motion), and Lorentz forces (induced by the magnetic field), all further modulated by the specific orientation of the cylinder. The following sections will analyze these interactions in detail.

\begin{figure}[h]
	\centering
	\includegraphics[width=1\textwidth]{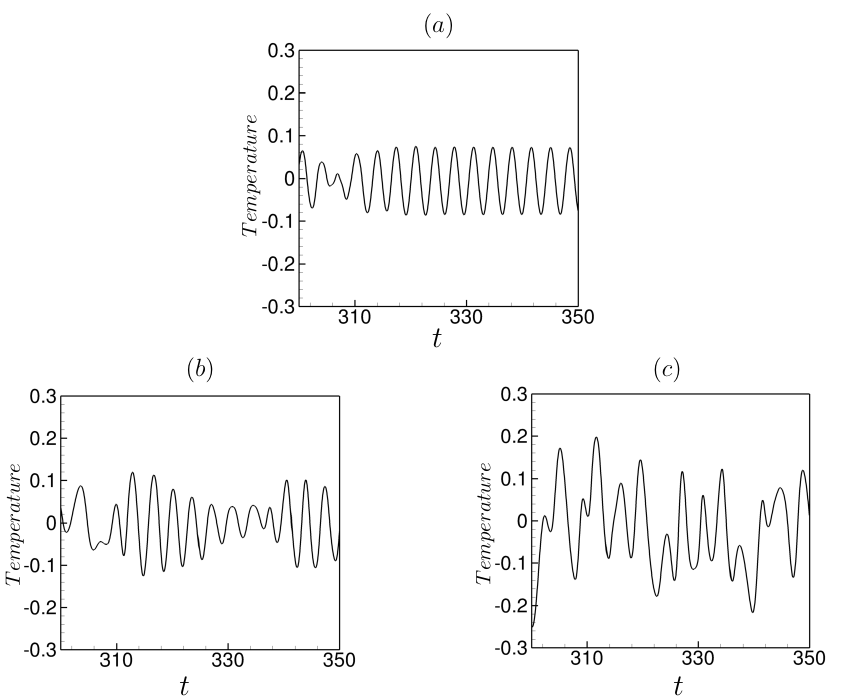}
	\caption{Temperature time history data for orientation 1 at a probe located $3.5h$ from prisms centroid at $Ha=0$ with varying $Ri$ : (a) $Ri=0$,(b) $Ri=1$ and (c) $Ri=5$ }
	\label{Temperature fluctuations}
\end{figure}
\begin{figure}[h]
	\centering
	% \begin{subfigure}
	%     \centering
	%      \hspace{0.6cm}{\large ($a$)} \hspace{0.5cm} \\
	%     \includegraphics[width=0.4\textwidth]{Fig_21a.eps} \\
	% \end{subfigure} 
	% \begin{subfigure}
	%     \centering
	%     \hspace{0.7cm}{\large ($b$)} \hspace{4.5cm} {\large ($c$)} \\
	%     \includegraphics[width=0.4\textwidth]{Fig_21b.eps}
	%    % \label{fig (b): Ux at x=1}
	% \end{subfigure}
	% \begin{subfigure}
	%     \centering
	%     \includegraphics[width=0.4\textwidth]{Fig_21c.eps}\\
	%    % \label{fig (b): Ux at x=1}
	% \end{subfigure}
	%\\
	% {\large ($a$)} \hspace{.5cm} {\large ($b$)} \hspace{2.5cm} {\large ($c$)}\\
	\includegraphics[width=1\textwidth]{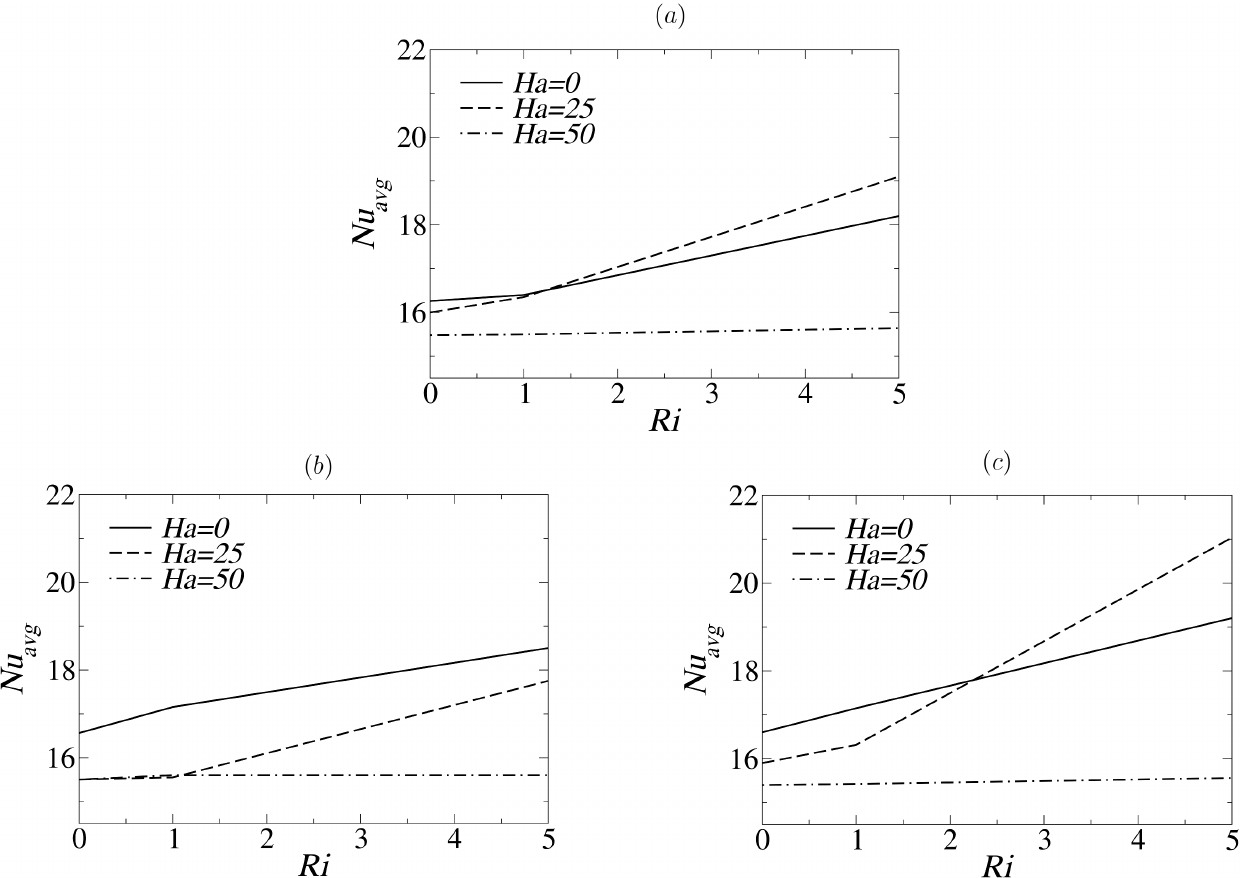}
	\caption{Variation of time-averaged mean Nusselt number over the bottom hot wall with $Ha$ and $Ri$ for all the three orientations: (a) Orientation 1, (b) Orientation 2, and (c) Orientation 3 }
	\label{Nu_vs_Ri}        
\end{figure}

The Richardson number ($Ri$) appears to play a significant role in heat transfer behavior for all cylinder Orientations. Temperature fluctuations measured at a distance of $3.5h$ downstream from the prism's centroid (Figure~\ref{Temperature fluctuations}) for Orientation 1 illustrate this effect. With $Ha = 0$ and $Ri = 0$, the temperature oscillates between $-0.08$ to $0.08$, indicating minimal temperature variation. However, as $Ri$ increases to $5$ (higher buoyancy forces), these fluctuations become more pronounced, expanding to a range of $-0.25$ to $0.2$. This signifies a substantial increase in temperature variation, reflecting stronger mixing due to buoyancy effects. Figure~\ref{Nu_vs_Ri} shows the time-averaged Nusselt number ($Nu_{avg}$) calculated over the entire bottom plate area. The local Nusselt number ($Nu_{local}$), the spatially averaged Nusselt number ($Nu$), and the time-averaged Nusselt number ($Nu_{avg}$)in the present study are given by :
\begin{equation*}
	Nu_{local} = \frac{L}{(T_h - T_c)}\frac{\partial T}{\partial y},\hspace{0.3cm}
	Nu = \frac{1}{A} \int Nu_{local} \hspace{0.1cm} dA, \hspace{0.3cm}
	Nu_{avg} = \frac{1}{\tau} \int Nu \hspace{0.1cm} dt
\end{equation*}
where $T_c$ and $T_h$ are the ambient fluid and hot wall temperatures, respectively. The time-averaged Nusselt number ($Nu_{avg}$) of the heated bottom wall exhibits a notable rise with increasing $Ri$ (Figure~\ref{Nu_vs_Ri}). The highest $Nu_{avg}$ values are observed for $Ri = 5$. This behavior can be attributed to the enhanced vertical fluid motion caused by stronger buoyancy forces (higher $Ri$). This intensified mixing within the wake region leads to more efficient heat transfer from the hot wall to the bulk flow. %Similar observations were reported by \cite{chatterjee2013wall} in their study on two-dimensional MHD flow over a circular cylinder. They found that an increase in wall vorticity strength corresponded to a rise in the Nusselt number.
In the present study, increasing $Ri$ also leads to enhanced vorticity near the heated wall, promoting heat transfer.

An interesting interplay between the Richardson number ($Ri$) and the applied magnetic field strength ($Ha$) is observed in the Nusselt number ($Nu$) behavior (Figure~\ref{Nu_vs_Ri}). At $Ri = 0~\&~1$ (weaker buoyancy effects), $Nu$ decreases for all orientations with increasing $Ha$ (Figure~\ref{Nu_vs_Ri}). This can be attributed to the Lorentz force induced by the magnetic field. The Lorentz force acts in opposition to the fluid inertia, weakening the vortex-shedding process behind the cylinder. Consequently, weaker vortex shedding leads to less efficient mixing and reduced heat transfer from the heated wall.

\begin{figure}[h]
	\centering
	\includegraphics[width=1\textwidth]{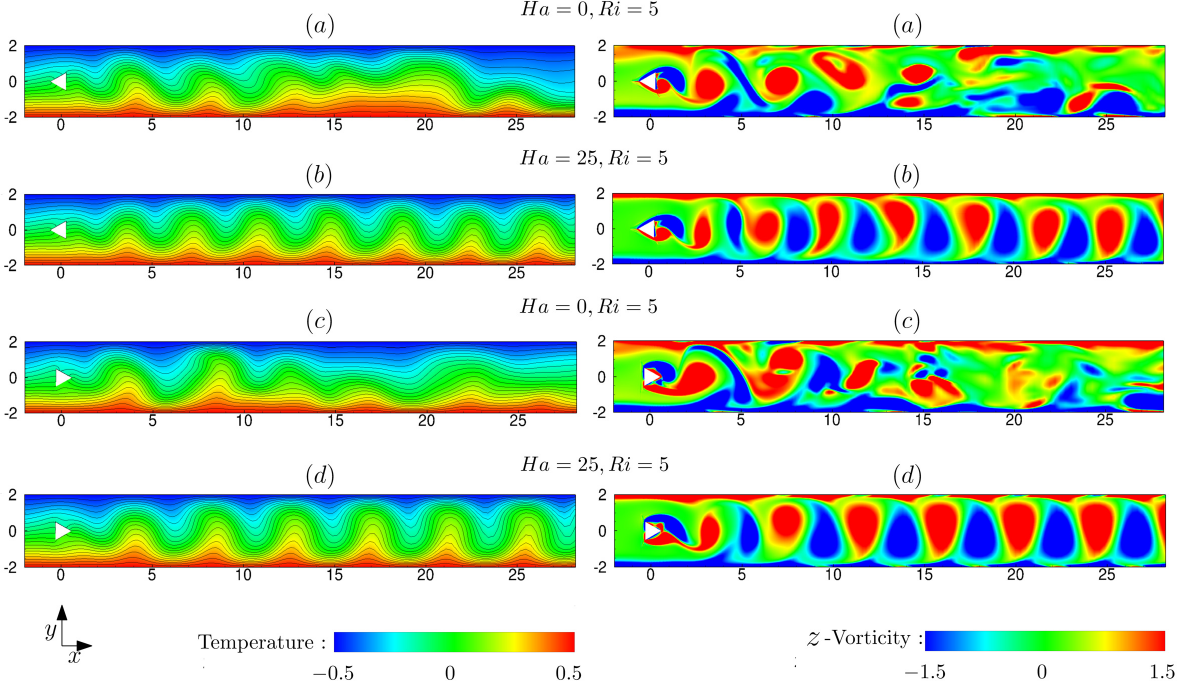}
	\caption{The figure shows isothermal (left) and vorticity (right) plots at constant $Ri$ of 5 for orientations 1 and 3: Figures $(a)$ and $(b)$ correspond to orientation 1 at $Ha=0$ and $Ha=25$ respectively, Figures $(c)$ and $(d)$ correspond to orientation 3 at $Ha=0$ and $Ha=25$ respectively}
	\label{temp_vort}
\end{figure}

However, a distinct trend emerges at $Ri = 5$ (stronger buoyancy effects) for Orientations 1 and 3. Here, $Nu$ is higher at $Ha = 25$ compared to $Ha = 0$ (Figure~\ref{Nu_vs_Ri} (a) and (c)). To understand this behavior, we need to consider the interaction between the buoyancy-driven flow and vortex shedding. At $Ha = 0$ and $Ri = 5$, the strong buoyancy effect generates a flow from the heated channel wall. This flow interacts with the shed vortices behind the cylinder, potentially leading to the breakdown of the regular vortex shedding pattern (refer to Figure~\ref{temp_vort} for illustration). This disruption results in a more chaotic flow field downstream, hindering efficient heat transfer. When $Ha$ increases to $25$ (stronger Lorentz force), the opposing force on the fluid inertia becomes more significant. This effectively reduces the local Reynolds number, promoting a more regular and persistent vortex shedding pattern all the way to the domain exit (Figure~\ref{temp_vort} $(b)$ ~\&~ $(d)$ for $Ha = 25~\&~Ri = 5$). This re-establishment of regular vortex shedding in Orientations 1 and 3 for $Ha = 25$ and $Ri = 5$ leads to enhanced mixing compared to the case with $Ha = 0$ and $Ri = 5$. The presence of this regular shedding process throughout the domain allows for better mixing of the flow, ultimately improving heat transfer from the heated wall. Further increasing $Ha$ to $50$ suppresses vortex shedding for all $Ri$ and Orientations, leading to a universal decrease in $Nu$ due to hindered heat transfer. Moreover, the Nusselt number remains constant for all $Ri$ and Orientations at $Ha=50$, indicating that the heat transfer behavior is least affected by the Orientation and $Ri$ value when the applied magnetic field strength is proportionally high.

\begin{figure}[h!]
	\centering
	\includegraphics[width=0.45\textwidth]{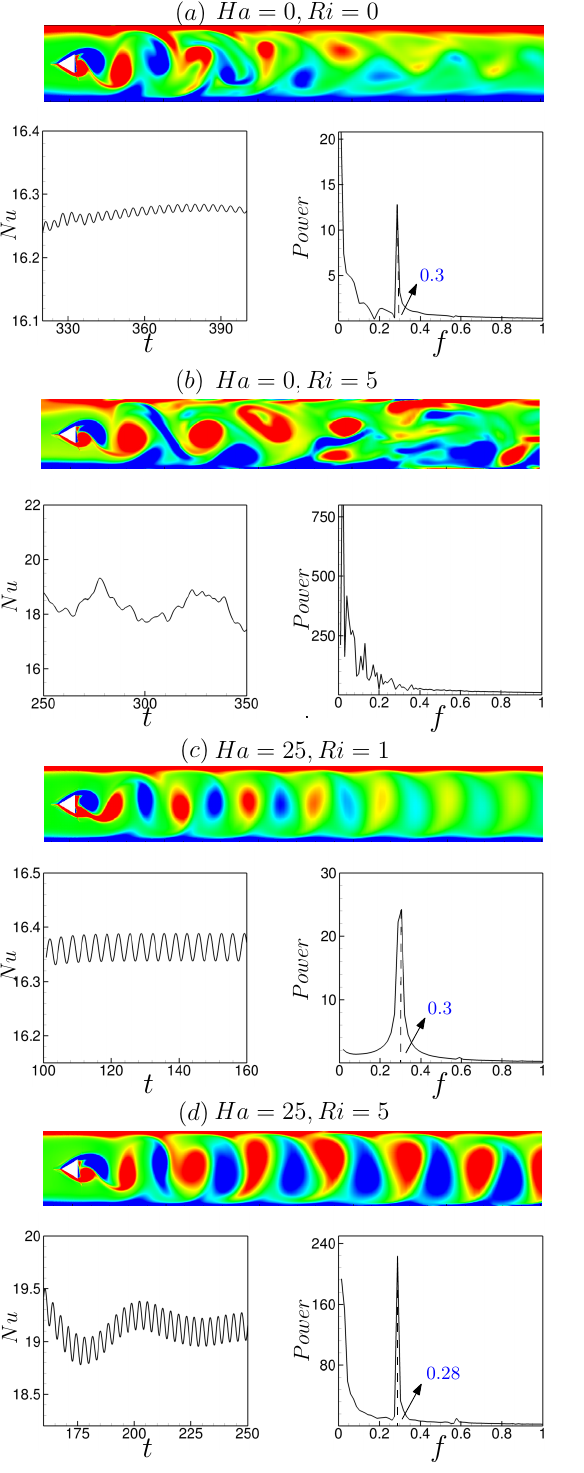}
	\caption{Time history of spatial mean Nusselt number over the bottom hot wall and its corresponding frequencies along with $z$ vorticity contours (plotted on top) for orientation 1: $(a)$ and $(b)$ correspond to $Ha=0, Ri=0$; $(c)$ and $(d)$ to $Ha=0, Ri=5$; $(e)$ and $(f)$ to $Ha=25, Ri=0$; $(g)$ and $(h)$ to $Ha=25, Ri=0$}
	\label{Nu_fft}
\end{figure}

Orientation 2 exhibits a distinct behavior in the interplay between $Ha$ and $Ri$ compared to Orientations 1 and 3 (Figure~\ref{Nu_vs_Ri} (b)). Here, $Nu$ generally decreases with increasing $Ha$ for all $Ri$ values, except for $Ha = 25~\&~50$ at $ Ri = 0~\&~1$. This exception can be attributed to the near or complete suppression of vortex shedding at $Ha = 25~\&~50$ for $ Ri = 0~\&~1$. With minimal/no vortex shedding, the steady flow dominates the heat transfer mechanism, leading to similar $Nu$ values in these cases.

The significant impact of $Ri$ on vortex shedding behavior and its consequent influence on heat transfer characteristics is further confirmed by the spatially averaged time-series data of the Nusselt number ($Nu$) (Figure~\ref{Nu_fft}). In mixed convection flows, where both forced convection (due to imposed flow) and free convection (due to buoyancy) coexist, the dynamics are significantly influenced by the interaction between the vortex shedding pattern and the thermal boundary layer. The periodic generation and shedding of vortices disrupt the boundary layer, causing it to thin and thicken periodically. This leads to fluctuations in heat transfer, reflected in the $Nu$ data. Consequently, fluctuations in lift coefficient (related to vortex shedding) and $Nu$ tend to be closely linked. For forced convection cases ($Ri = 0$ at all $Ha$ values), the dominant frequency obtained from the time-series data of the spatially averaged $Nu$ over the heated bottom plate matches the dominant frequency calculated from the lift coefficient data (Figure~\ref{Nu_fft} ($a$)). This is because, at $Ri = 0$, vortex shedding is the primary mechanism for heat transfer from the bottom wall. However, when buoyancy effects become dominant ($Ri > 0$), heat transfer is no longer solely governed by shedding behavior. The interaction of the heated wall flow with the vortex street modulates the frequencies, suggesting a more complex and potentially chaotic flow structure (Figure~\ref{Nu_fft} ($b$)). It is also noted that when the flow behavior is two-dimensional, the dominant frequencies obtained from the lift coefficient and $Nu$ data align, regardless of the cylinder orientation (Figure~\ref{Nu_fft} ($c$)). Conversely, discrepancies arise between the dominant frequencies extracted from lift coefficient and $Nu$ data in scenarios where the flow exhibits significant three-dimensional characteristics, highlighting the importance of flow dimensionality in the relationship between vortex shedding and heat transfer.

The present study highlights the intricate interplay between obstacle orientation (1, 2, and 3), buoyancy effects (characterized by $Ri$), and magnetic field strength $(Ha$) on heat transfer behavior. These variables influence the interaction between the channel wall boundary layer vortices and the vortex street shed behind the obstacle, leading to diverse wake dynamics for different parameter combinations. These wake dynamics ultimately dictate the heat transfer characteristics. Across all orientations, Orientation 3 exhibited the highest heat transfer rate at $Ha = 25~\&~Ri = 5$, with an average Nusselt number of $Nu_{avg}=21.05$. This suggests a more efficient mixing process within the wake region for this specific orientation under these conditions. In comparison, Orientations 1 and 2 displayed lower average $Nu$ values of $19.1$ and $17.75$, respectively, at $Ha = 25~\&~Ri = 5$. Moreover, Orientation 3 consistently demonstrated a higher heat transfer rate than the other two orientations for various $Ha$ and $Ri$ combinations. These observations emphasize the importance of considering the obstacle orientation when designing heat transfer systems involving bluff bodies.

\section{Conclusion \label{sec:concl}}
The study provides a detailed investigation of three-dimensional wake dynamics and heat transfer characteristics in MHD flows over triangular cylinders confined between parallel plates. By considering varying Hartmann numbers ($Ha=0,25,50$) and Richardson numbers ($Ri=0,1,5$) at a fixed Reynolds number ($Re_{h} =600$), the work addresses critical knowledge gaps in understanding MHD flows at low Hartmann numbers, focusing on the interplay between magnetic fields, buoyancy effects, and geometric orientation.
\begin{itemize}
	\item Wake Instabilities: Three distinct wake instabilities, Modes A, B, and C were identified, varying in transverse vorticity structure. At $Ha=0$, parallel vortex streets dominated, while $Ha=25$ induced regular alternate vortex shedding patterns. An intermittent delay in vortex shedding is observed at higher $Ri$ for Orientations 1 and 3 of the triangular cylinder.
	\item Flow Separation and Recirculation: The mean recirculation length ($L_r$) showed a strong correlation with $Ha$ and $Ri$, displaying a critical $Ha$ beyond which $L_r$ decreased. In Orientations 1 and 3, $L_r$ exhibited a notable drop during transitions from three-dimensional to two-dimensional wake.
	\item Drag and Lift Characteristics: Orientation 3, with its wider wake, exhibited the highest drag coefficient ($C_d$) and the lowest vortex shedding frequency. The mean drag coefficient ($\overline{C_d}$) increased with $Ha$ and decreased with $Ri$ across all orientations. The symmetric geometry of Orientations 1 and 3 resulted in a zero mean lift coefficient ($\overline{C_l}$) for $Ri = 0$. Conversely, the asymmetric geometry of Orientation 2 led to non-zero mean $\overline{C_l}$ due to unequal size vortex shedding from either side of the cylinder. Here, the lift force magnitude decreased with increasing $Ha$.
	\item Heat Transfer Performance: Heat transfer performance was strongly influenced by $Ri$, with buoyancy-enhanced convection driving higher Nusselt numbers. The highest heat transfer rate ($Nu_{avg} = 21.05$) was observed for Orientation 3 at $Ha = 25~\&~ Ri = 5$. The dominant frequency from the time-series data of the Nusselt number ($Nu$) matched the dominant frequency from the time-series data of the lift coefficient ($C_l$) in forced convection flows. In mixed convection, this alignment was observed in two-dimensional flows; however, discrepancies appeared in flows exhibiting significant three-dimensionality. 
\end{itemize}

\backmatter

\bmhead{Acknowledgements}

The authors would like to thank the Indian Institute of Technology Hyderabad and the Government of India for providing the High-Performance Computational facility PARAM Seva under the National Supercomputing Mission for carrying out the simulations.

\bmhead{Author Contributions}
Amulya Sai Akkaladevi contributed methodology, software, validation, formal analysis, investigation, writing (original draft;review and editing), and visualization. Prabhat Kumar contributed writing (review and editing) and supervision. Sachidananda Behera contributed conceptualization, methodology, resources, writing (review and editing), supervision and project administration.

\bmhead{Data Availability}
The data that support the findings of this study are available from the corresponding author upon reasonable request.

\section*{Declarations}

\textbf {Competing interests} The authors declare no competing interests.

\noindent

%%===========================================================================================%%
%% If you are submitting to one of the Nature Portfolio journals, using the eJP submission   %%
%% system, please include the references within the manuscript file itself. You may do this  %%
%% by copying the reference list from your .bbl file, paste it into the main manuscript .tex %%
%% file, and delete the associated \verb+\bibliography+ commands.                            %%
%%===========================================================================================%%

\bibliography{ref}% common bib file
%% if required, the content of .bbl file can be included here once bbl is generated
%%\input sn-article.bbl

\end{document}